\documentclass[10pt]{amsart}

\usepackage{amssymb}
\usepackage{latexsym}
\usepackage{amsmath}
\usepackage{esint}
\usepackage{amsthm}
\usepackage{soul}
\usepackage{xcolor}
\usepackage{tikz}
\usepackage{float}
\usetikzlibrary{patterns, decorations.pathreplacing}
\usepackage{hyperref} 
\hypersetup{
    colorlinks=true,    
    linkcolor=blue,
    citecolor=blue,
    urlcolor=blue}
\usepackage{bbold} 
\usepackage{enumitem} 
\usepackage{thmtools} 
\usepackage[cal=euler]{mathalpha} 

\makeatletter
\def\@email#1#2{%
 \endgroup
 \patchcmd{\titleblock@produce}
  {\frontmatter@RRAPformat}
  {\frontmatter@RRAPformat{\produce@RRAP{*#1\href{mailto:#2}{#2}}}\frontmatter@RRAPformat}
  {}{}
}%
\makeatother

\definecolor{fgreen}{RGB}{34,169,75}
\definecolor{gpurple}{RGB}{152,68,158}

\newtheorem{theorem}{Theorem}[section]
\newtheorem{corollary}[theorem]{Corollary}
\newtheorem{lemma}[theorem]{Lemma}

\newcounter{definition}
\newenvironment{definition}{\vspace{0.7\baselineskip} \refstepcounter{definition} \par \noindent \textbf{Definition~\thedefinition.}}{\vspace{0.7\baselineskip}}
\numberwithin{definition}{section}

\newcounter{example}
\newenvironment{example}{\vspace{0.7\baselineskip} \refstepcounter{example} \par \noindent \textbf{Example~\theexample.}}{\vspace{0.7\baselineskip}}
\numberwithin{example}{section}

\newcommand{\R}{\mathbb{R}}

\newcommand{\Z}{\mathbb{Z}}
\newcommand{\N}{\mathbb{N}}
\newcommand{\E}{\mathbb{E}}
\renewcommand{\P}{\mathbb{P}}

\renewcommand{\epsilon}{\varepsilon}

\newcommand{\word}[1]{\quad \text{#1} \quad}

\newcommand{\paren}[1]{\left( #1 \right)}
\newcommand{\bracket}[1]{\left[ #1 \right]}
\newcommand{\cparen}[1]{\left\{ #1 \right\}}

\newcommand{\abs}[1]{\left| #1 \right|}
\newcommand{\norm}[1]{\left\| #1 \right\|}
\newcommand{\var}{\mathrm{Var}}

\newcommand{\truncated}[2][k]{\mathrm{max}\{\mathrm{min}\{#2, #1\}, -#1\}}

\allowdisplaybreaks

\renewcommand{\eqref}[1]{(\ref{#1})}

\newcommand\numberthis{\addtocounter{equation}{1}\tag{\theequation}}

\newcommand\blfootnote[1]{%
  \begingroup
  \renewcommand\thefootnote{}\footnote{#1}%
  \addtocounter{footnote}{-1}%
  \endgroup
}

\def\equationautorefname#1#2\null{%
  Equation #1(#2\null)%
}

\begin{document}

\title{Localization of the 1D Non-Stationary Anderson Model}

\author[K. Zieber]{Karl Zieber}
\date{\today}
\blfootnote{K.Z. was supported in part by NSF DMS Grant 2247966.}

\begin{abstract}
This paper considers the family of Schrödinger operators on $\ell^2(\Z)$ given by independent but not necessarily identically distributed and possibly unbounded potentials. We assume a finite exponential moment and allow the choice of distributions to come from any compact set away from deterministic distributions. With these assumptions we prove spectral localization with exponentially decaying eigenfunctions as well as dynamical localization. One of the main tools is a Furstenberg-type theorem for non-stationary matrix products.  
\end{abstract}

\maketitle


\section{Introduction}

The focus of this paper is the celebrated Anderson model, which is the Schrödinger operator on $\ell^2(\mathbb{Z})$ defined as
\begin{equation}\label{eq:Schrod}
\left[H \psi\right](n)=\psi(n+1)+\psi(n-1)+V(n) \psi(n)
\end{equation}
where the potentials $V(n)$ are chosen randomly and independently. This model was introduced by P. W. Anderson in 1958 to understand how randomness can impede the propagation of electrons in disordered media \cite{Anderson:1958aa}. When electrons (i.e., eigenfunctions) are ``trapped'' in a finite-volume region with sufficiently high probability, we say the operator experiences localization. Several notions of localization have been developed, including ``spectral localization'' and its stronger cousin ``dynamical localization.'' Since the introduction of the original model, physicists and mathematicians have studied numerous variants of the Anderson model and their properties. In addition to adjusting the dimension, analogous operators have been considered in continuous media and on lattices with structures other than $\mathbb{Z}^d$, see e.g. \cite{Aizenman:2015aa, Damanik:2020aa, Klein:1994aa, Klein:1990aa, Macera:2022aa}. 

The first result that mathematically established localization in the 1-dimensional case was proved by Goldsheid, Molchanov, Pastur \cite{Goldshtein:1977aa}, who considered the 1-dimensional continuous version of the Anderson model. Later, Kunz and Souillard considered the discrete setting in \cite{Kunz:1980aa}, proceeding by first proving dynamical localization and from there obtaining spectral localization.  Initially the methods developed were only applicable to suitably nice (e.g., absolutely continuous) distributions of the potentials $V(n)$. A significant step was taken by Carmona, Klein, Martinelli \cite{Carmona:1987aa} in their proof of localization for singular potentials using the fruitful multi-scale analysis (MSA). Notably, \cite{Carmona:1987aa} required only a mild finite-moment condition for the 1-dimensional case and utilized Furstenberg's theorem on the product of random matrices, an approach going back at least to 1985 (see \cite{Frohlich:1985aa}). Localization in the 1-dimensional, discrete, stationary case (where stationary means potentials $V(n)$ are iid) was settled by \cite{Carmona:1987aa}. Some improvements were made in 1993 when Aizenman and Molchanov developed the fractional moment method in \cite{Aizenman:1993aa}, which sidesteps some of the heavy machinery of MSA while still reaping many of the benefits. In particular, \cite{Aizenman:1993aa} requires absolutely continuous distributions for the random potentials but not stationarity nor boundedness.For a more detailed description of background, existing methods, or current results see \cite{Bucaj:2019aa, Carmona:1990aa, damanik2011, DAMANIK:2017aa, Damanik:2022aa, Damanik:2025aa, hislop08, Hundertmark07, stolz2011}.

Recently, Bucaj et al. presented a more straightforward version of the proof of localization in \cite{Bucaj:2019aa}, sparking renewed interest in the 1-dimensional discrete model. Shortly after, Jitomirskaya and Zhu presented a proof of localization throughout the spectrum \cite{Jitomirskaya:2019aa} using an approach that had been developed by Jitomirskaya to originally study almost-Mathieu operators \cite{Jitomirskaya:1999aa}. Around the same time, Gorodetski and Kleptsyn proved a parametric version of Furstenberg's theorem and applied it to the stationary 1D Anderson model \cite{Gorodetski:2021aa}, yielding spectral and dynamical localization via a different method from Jitomirskaya and Zhu. Later, Gorodetski and Kleptsyn extended this to the non-stationary setting. They proved a non-stationary analog of Furstenberg's theorem in \cite{Gorodetski:2023aa} and used this result to achieve spectral and dynamical localization in the non-stationary 1-dimensional Anderson model of the form \eqref{eq:Schrod} in \cite{Gorodetski:2024aa}. Additionally, localization for non-stationary potentials was established by Hurtado in \cite{Hurtado:2024aa} at the bottom of the spectrum for the 2-dimensional Anderson model on $\ell^2(\Z^2)$. However, both \cite{Gorodetski:2024aa} and \cite{Hurtado:2024aa} assume the distributions of the potentials $V(n)$ all have supports contained in a common compact interval. This paper removes the need for uniformly bounded potentials to conclude spectral localization for the 1D Anderson model as formulated in the following theorem:

\begin{theorem}\label{main_result}
Consider the random Schrödinger operator $H$ acting on $\ell^2(\mathbb{Z})$ via
\[
\left(H \psi\right)(n)=\psi(n-1)+\psi(n+1)+V(n) \psi(n)
\]
where the potentials $V(n)$ are chosen randomly and independently, each with distribution denoted by $\mu_n$. It is assumed that: 
\begin{itemize}
	\item \textbf{Finite $\gamma$-moment:} There exists a $\gamma >0$ and $C_0$ such that for any $n$ we have $\int|x|^\gamma d\mu_n(x) < C_0$.
	\item \textbf{No deterministic distributions:} There is a compact interval upon which all the distributions have some minimal amount of variation when restricted to that interval. In detail, there exists an $\epsilon >0$ and a $k > 0$ such that for any $n$ we have $\mathrm{Var}(\truncated{V(n)}) > \epsilon$.
\end{itemize}
Then, almost surely, $H$ is spectrally localized.
\end{theorem}

We wish to remark that in the case when $\gamma > 2$, the second hypothesis may be weakened slightly to only assuming that $\mathrm{Var}(V(n)) > \epsilon$, with no need for the variation to persist when considering the restriction to $[-k,k]$.
The difference between the requirements for the $\gamma > 2 $ and $\gamma \leq 2$ regimes is described in more detail in Appendix \ref{assumptions}. 

Furthermore, a stronger notion of localization also holds. We say a self-adjoint operator $H: \ell^2(\mathbb{Z}) \rightarrow \ell^2(\mathbb{Z})$ has dynamical localization if for any $q>0$ one has
\[
\sup _t \sum_{n \in \mathbb{Z}}(1+|n|)^q\left|\left\langle\delta_n, e^{-i t H} \delta_0\right\rangle\right|<\infty .
\]
\vspace{-\baselineskip}
\begin{theorem}\label{dynam_loc}
Under the assumptions of \autoref{main_result}, $H$ is dynamically localized.
\end{theorem}
In fact, we will show the stronger property that $H$ has semi-uniformly localized eigenfunctions (SULE). See Section \ref{Dynamical Localization} for more details.

The method used to establish these results relies on a non-stationary version of Furstenberg's Theorem, which was proved by Gorodetski and Kleptsyn in \cite{Gorodetski:2023aa} and used in their 2025 localization result \cite{Gorodetski:2024aa}. Much of the rest of the paper is inspired by the approach taken by Jitomirskaya and Zhu in \cite{Jitomirskaya:2019aa} and Rangamani in \cite{Rangamani:2019aa}. This approach was also used by \cite{Ge:2020aa} to achieve exponential dynamical localization in expectation for the one-dimensional Anderson model. Spectral localization with unbounded potentials has been addressed in the heavy-tailed stationary case with multi-scale analysis methods in \cite{Hurtado:2025aa}.

The rest of this paper is organized as follows. We pay attention primarily to proving spectral localization and then later adapting some of the lemmas to also conclude dynamical localization. In Section \ref{Preliminaries}, we discuss common tools derived from the properties of the truncated Schrödinger operator, including finite-volume Green's functions and their use in reducing the spectral localization claim in \autoref{main_result} to the more technical \autoref{main_theorem_reform}. Section \ref{Furstenbergs Theorem} contains the non-stationary Furstenberg theorem from \cite{Gorodetski:2023aa} and necessary corollaries. These results allow us to establish large deviations estimates not only for our transfer matrices but also their entries, a technical but important result in the spirit of \cite{tsay99, Bucaj:2019aa}. In Section \ref{Properties}, we use these large deviations estimates to establish equicontinuity of the functions which give the ``typical'' behavior of our matrix products. One of the challenges of the non-stationary setting is the loss of the Lyapunov exponent, but equicontinuity of this family of functions acts as an analog of the continuity of the Lyapunov exponent. Following this, in Section \ref{Technical Lemmas} we prove estimates for the sets where large deviations do occur. In Section \ref{Proof of Main Theorem} we prove \autoref{main_theorem_reform} (spectral localization) and Section \ref{Dynamical Localization} contains the proof of dynamical localization. We discuss the necessity of the ``no deterministic distribution'' assumptions under the different $\gamma$-moment regimes in Appendix \ref{assumptions} and provide an example of a sequence of distributions for which localization was previously unknown but now holds via \autoref{main_result} in Appendix \ref{example_covered}. 


\section{Standard Spectral Tools}\label{Preliminaries}

The goal of this section is to introduce some standard tools used to study the 1-dimensional discrete Anderson operator. The two objects of particular interest to us are the Green's functions and the characteristic polynomials associated to the finite-volume, truncated version of the operator $H$. These have a close relationship that we can exploit (see \eqref{G_to_P_eq}) and are also closely related to the ``transfer matrices'' that we discuss in Section \ref{Furstenbergs Theorem}. We will also discuss how we are able to reduce the main result, \autoref{main_result}, to \autoref{main_theorem_reform} using previously-established results and techniques.

Recall that we denote the distribution of $V(n)$ as $\mu_n$. Set $\P := \prod_{n\in \Z} \mu_n$ to be the full probability measure on $\Omega:=\R^\Z$ unless otherwise stated. For the remainder of the paper, we will consider the family of Schrödinger operators $H_\omega: \ell^2(\mathbb{Z}) \to \ell^2(\mathbb{Z})$ defined as
\begin{equation}
\left[H_\omega \psi\right](n)=\psi(n+1)+\psi(n-1)+V_\omega(n) \psi(n),
\end{equation}
where $\omega \in \Omega$ and $V_\omega(n)$ is a particular realization of $V(n)$. Note that, for any $\omega$, $H_\omega$ is self-adjoint and hence has only real eigenvalues. We say $H_\omega$ experiences \textit{spectral localization} with exponentially decaying eigenfunctions (sometimes referred to as \textit{Anderson localization}) when two things hold for the operator $H_\omega$:
\begin{enumerate}
	\item The spectrum of $H_\omega$ is $\P$-almost surely ``pure point.'' We say that $H_\omega$ has $\mathbb{P}$-almost surely pure point spectrum if for $\mathbb{P}$-almost every $\omega \in \Omega$ and every $\psi \in \ell^2(\mathbb{Z})$ the spectral measure $\mu_{\psi}$ (where $\mu_{\psi}$ is such that $\langle \psi , f(H) \psi \rangle = \int_{\sigma(H_\omega)} f(E) d\mu_{\psi}(E)$ for all $f \in C(\sigma(H_\omega))$) is a pure point measure.
	\item The eigenfunctions $\psi$ (corresponding to eigenvalues $E$ which are in some suitable range) are exponentially decaying, i.e., $|\psi(n)|$ decays at least exponentially as $n \to \pm\infty$. 
\end{enumerate}
For more technical details about spectral localization see \cite{Carmona:1990aa, Damanik:2022aa, Damanik:2025aa}. In our case, Sch'nol's Theorem (see \cite{Shn57, Han:2019aa}) guarantees the spectrum is almost surely pure-point, provided that we can show all polynomially-bounded solutions to $H_\omega \psi = E \psi$ decay at least exponentially fast. This leads to the following definition:

\begin{definition}
We call an energy $E \in \R$ a \textit{generalized eigenvalue} if there exists a $\psi \in \R^\Z$ which is polynomially bounded and we have $H_\omega \psi = E \psi$. In this case we call $\psi$ a \textit{generalized eigenfunction}.
\end{definition}

With this in mind, we may prove \autoref{main_result} by proving the following theorem:

\begin{theorem}\label{main_theorem}
Under the assumptions of \autoref{main_result}, for a.e. $\omega$, for every generalized eigenvalue $E$, the corresponding generalized eigenfunction $\psi_{\omega, E}(n)$ decays exponentially as $|n|\to \infty$.
\end{theorem}

To establish this, we leverage the well-studied tools of the truncated Schrödinger operator. We provide the necessary background and refer the reader to \cite{Damanik:2022aa} Section 2.2 for more details. Let $a, b \in \Z$ and let $H_{[a, b], \omega}$ denote the Schrödinger operator restricted to the interval $[a,b]\cap \Z$ with Dirichlet boundary condition. Letting $\sigma(H_{[a,b], \omega})$ denote the spectrum of $H_{[a, b], \omega}$, we define the \textit{Green's function} associated to $E \not\in \sigma(H_{[a,b], \omega})$ as
\[G_{[a,b], E, \omega} := (H_{[a,b], \omega} - E)^{-1}.\]
Since 
\[H_{[a,b], \omega} - E = \begin{bmatrix}
V_\omega(a) - E & 1 & & & 0\\
1 & V_\omega(a+1) - E & 1 & &\\
& 1 &  & \ddots & \\
& & \ddots & & 1 \\
0 & & & 1 & V_\omega(b) - E \\
\end{bmatrix},
\]
we have that $G_{[a,b], E, \omega}$ is a $(b-a + 1)$-dimensional matrix as well. For a matrix $A$, we write the $(j,k)$-entry of $A$ as $A(j,k)$. It is known (see \cite{Damanik:2022aa} Chapter 2, Section 2) that if $\psi$ solves $H_\omega \psi = E \psi$ then
\begin{equation}\label{eq:green-to-ev}
\psi(x)=-G_{[a, b], E, \omega}(x, a) \psi(a-1)-G_{[a, b], E, \omega}(x, b) \psi(b+1), \quad x \in[a, b].
\end{equation}

This equation allows us to control the behavior of a generalized eigenfunction at a particular point by using the Green's function and select endpoints. Thus our strategy is to control the Green's function at suitable points to demonstrate exponential decay. As such, we define:

\begin{definition}
For $C>0,$ $n \in \mathbb{N}$, we say $x \in \mathbb{Z}$ is $(C, n, E, \omega)$\textit{-regular}, if
\[\abs{G_{[x-n, x+n], E, \omega}(x, x-n)} \leqslant e^{-C n} \word{and} \abs{G_{[x-n, x+n], E, \omega}(x, x+n)} \leqslant e^{-C n}.\]
Otherwise, we call it $(C, n, E, \omega)$\textit{-singular}.
\end{definition}

With this, we may reformulate \autoref{main_theorem}:

\begin{theorem}\label{main_theorem_reform}
For almost every $\omega$, for every generalized eigenvalue $E$ of $H_\omega$, there is some $C>0$ and $N(\omega, E)$ such that for every $n>N$, $2 n$ and $2 n+ 1$ are $(C, n, E, \omega)$-regular.
\end{theorem}

To see how this implies \autoref{main_theorem}, suppose for all $n$ sufficiently large, $2 n+ 1$ is $(C, n, E, \omega)$-regular. Then, for the generalized eigenfunction $\psi$ with generalized eigenvalue $E$, we invoke \eqref{eq:green-to-ev}:
\begin{align*}
	|\psi(2n+1)| & \leq \abs{G_{[n+1, 3n+1],E,\omega}(2n+1, (2n+1) - n) \psi(n)} ]\\
	& \qquad + \abs{G_{[n+1, 3n+1],E,\omega}(2n+1,(2n+1) + n) \psi(3n+2)} \\
	& \leq  e^{-Cn}\cdot |\psi(n)| + e^{-Cn} \cdot |\psi(3n+2)|.
\end{align*}
Since $\psi$ is polynomially bounded, this gives exponential decay of $|\psi(2n+1)|$ as $n \to \infty$. The argument for $2n$ is identical. To get decay as $n \to -\infty$, showing $2 n+ 1$ is $(C, |n|, E, \omega)$-regular is also an identical argument.


\section{Furstenberg's Theorem and Large Deviation Estimates}\label{Furstenbergs Theorem}

In this section, we will establish the tools from random matrix theory that will play a crucial role in our estimates of the Green's functions. Most important is \autoref{noniidFurst} which gives us a large deviations estimate for our random matrices. This is used to establish a similar deviations result for the characteristic polynomials above as well as equicontinuity of a critical family of functions in Section \ref{Properties}.

For the remainder of this paper, we consider energies in an arbitrary, fixed, compact interval $E \in J$. We set 
\[A_{k, E, \omega} = \begin{bmatrix}
E - V_\omega(k) & -1\\
1 & 0\\
\end{bmatrix}
\]
as our \textit{transfer matrices}. Let $T_{[a,b], E, \omega} = \prod_{k=a}^{b} A_{k, E, \omega}$. If $\psi$ solves $H_\omega \psi = E \psi$, then entries of $\psi$ are determined by two initial points and products of $A_{k, E, \omega}$. Indeed,
\begin{equation}\label{transfer_eq}
\begin{bmatrix}
\psi(b+1)\\
\psi(b)\\
\end{bmatrix} = T_{[a,b], E, \omega}\begin{bmatrix}
\psi(a)\\
\psi(a-1)\\
\end{bmatrix}.
\end{equation}
For ease of notation, let $T_{n, E, \omega} = \prod_{k=1}^{n} A_{k, E, \omega}$. Denote $L_{n, E} = \E[\log\norm{T_{n, E, \omega}}]$ where the expectation is taken over the distribution $\mu^E_1 \times \mu^E_2 \times \ldots \times \mu^E_n$, the push-forward distributions for $A_{k, E, \omega}$ on $\mathrm{SL}(2, \R)$ derived from the original distributions for the potentials. Similarly denote $L_{[a,b], E, \omega} = \E[\log\norm{T_{[a,b], E, \omega}}]$, which will be colloquially referred to as ``growth functions.'' One way to think of these growth functions is as tracking the average (i.e., typical) growth behavior of our transfer matrix norms. We wish to establish a large deviations estimate for $\log\norm{T_{n, E, \omega}}$ and we will use these growth functions to do so.

To establish these large deviation estimates, we invoke a theorem from \cite{Gorodetski:2024aa} which has some technical requirements. Let us check these requirements are satisfied in our context. For each $A \in \mathrm{SL(2, \R)}$, let $f_A :\mathbb{S}^1 \to \mathbb{S}^1 $ denote the projectivization of $A$. Now, our assumptions in \autoref{main_result} imply:
\begin{enumerate}
	\item[\textbf{(A1)}] \label{list:assumption1} All $\mu^E_n$ belong to some weak$^*$-compact set of measures $\mathcal{K}$.
	\item[\textbf{(A2)}] \label{list:assumption2} There exists $\gamma>0, C$ such that for every $E \in J$ and every $\mu^E \in \mathcal{K}$:
    $$
    \int_{\mathrm{SL}(2, \mathbb{R})}\|A\|^\gamma d \mu^E(A)<C
    .$$
    \item[\textbf{(A3)}] For any $E \in J$ and any measure $\mu^E \in \mathcal{K}$ there are no Borel probability measures $\nu_1, \nu_2$ on $\mathbb{R P}^1$ such that $\left(f_A\right) * \nu_1=\nu_2$ for $\mu^E$-almost every matrix $A \in \mathrm{SL}(2, \mathbb{R})$.
\end{enumerate} 
Confirming these statements hold in our context, first note the finite $\gamma$-moment assumption in \autoref{main_result} implies \textbf{(A2)}. 

Statement \textbf{(A3)} follows from the form of our transfer matrices and technically holds for any \textit{pair} of transfer matrices from $\mathrm{SL}(2, \mathbb{R})$ (see \cite{Gorodetski:2024aa}, Section 4). Suppose we have two Borel probability measures on $\mathbb{R P}^1$ such that $f * \nu_1= g * \nu_1 = \nu_2$, where $f$ and $g$ are projectivizations of two different transfer matrices. Then $(f \circ g^{-1}) * \nu_2 = \nu_2$. However, notice
\[A_{k, E, \omega_1} \cdot A_{k, E, \omega_2}^{-1} = \begin{bmatrix}
1 & *\\
0 & 1\\
\end{bmatrix}
\]
And so $\nu_2$ must be the probability measure with point mass in the direction of $[1, 0]^T$, i.e., $\delta_{[1, 0]^T}$. However, for any transfer matrix $A$, $f_A * \nu_1 = \delta_{[1, 0]^T}$ implies $\nu_1 = \delta_{[0, -1]^T}$. Since $\nu_1 \neq \nu_2$, under the composition of two transfer matrices there is no measure with a deterministic image and \textbf{(A3)} is satisfied. 

For \textbf{(A1)}, we care about compactness in the weak$^*$ sense, but we will actually argue sequential compactness in the weak topology\footnote{Here, we say $\mu_n \to \mu$ in the weak$^*$ topology if $\int fd\mu_n \to \int fd\mu$ for every continuous function $f:\R \to \R$ vanishing at $\infty$. We say $\mu_n \to \mu$ in the weak topology if $\int fd\mu_n \to \int fd\mu$ for every continuous, bounded function $f:\R \to \R$.}, which is sufficient. Since the distributions $\mu_n^E$ on $\mathrm{SL}(2, \R)$ are related to the distributions of the potentials $\mu_n$ on $\R$ by a continuous push-forward map, it is enough to establish that $\{\mu_n\}_{n \in \Z}$ is compact in the weak sense by the Continuous Mapping Theorem. Prokhorov's theorem (see \cite{Billingsley:1999aa} Chapter 1, Section 5) allows us to claim weak compactness provided we can show this family of probability measures is tight. Here, we say a family of measures is \emph{tight} if for every $\epsilon > 0$ there exists a compact set $K \subseteq \R$ such that $\mu_n(K) > 1 - \epsilon$ for all $n$. Indeed, by Markov's inequality we have
\[\mu_n(\{|x| > R\}) \leq \frac{1}{R^\gamma} \int |x|^\gamma d\mu_n < \frac{C}{R^\gamma}\]
for any $R >0$. Setting $\epsilon > 0$, choose $R_\epsilon \gg 1$ so that $\frac{C}{R^\gamma_\epsilon} < \epsilon$. Denoting $K_\epsilon = \cparen{|x| \leq R_\epsilon}$ and noting this is a compact set, we have
\[\mu_n(K_\epsilon) > 1 - \frac{C}{R^\gamma_\epsilon} > 1 - \epsilon\]
for any $n$, which establishes tightness. Hence statement \textbf{(A1)} holds for our context.

These statements \textbf{(A1)}-\textbf{(A3)} allow us to invoke the following theorem of Gorodetski and Kleptsyn. We state the result for $\mathrm{SL}(2, \R)$ but note that the original theorem holds for $\mathrm{SL}(d, \R)$.

\begin{theorem}[Non-Stationary Furstenberg Theorem and Large Deviations Estimates, Theorem 2.1 in \cite{Gorodetski:2024aa}] \label{noniidFurst}
Let $\mathcal{K}$ be a compact set of probability measures on $\mathrm{SL}(2,\R)$. Assume that the following hold:
\begin{itemize}
	\item \textbf{(finite moment condition)} There exists $\gamma>0, C$ such that
    $$
    \forall \mu \in \mathcal{K} \quad \int_{\mathrm{SL}(2, \mathbb{R})}\|A\|^\gamma d \mu(A)<C.
    $$
    \item \textbf{(measures condition)} For any $\mu \in \mathcal{K}$ there are no Borel probability measures $\nu_1, \nu_2$ on $\mathbb{R P}^{1}$ such that $\left(f_A\right)_* \nu_1=\nu_2$ for $\mu$-almost every $A \in \mathrm{SL}(2, \mathbb{R})$.
\end{itemize}
Then for any $\varepsilon>0$ there exists $\delta>0$ such that for any sequence of distributions $\mu_1, \mu_2, \ldots, \mu_n, \ldots$ from $\mathcal{K}$, for all sufficiently large $n \in \mathbb{N}$ we have
$$
\mathbb{P}\left\{\left|\log \left\|T_n\right\|-L_n\right|>\varepsilon n\right\}<e^{-\delta n},
$$
where $T_n=A_n A_{n-1} \ldots A_1,\left\{A_j\right\}$ are chosen randomly and independently with respect to $\left\{\mu_j\right\}, \mathbb{P}=\mu_1 \times \mu_2 \times \ldots \times \mu_n$, and $L_n=\mathbb{E}\left(\log \left\|T_n\right\|\right)$. Moreover, the same estimate holds for the lengths of random images of any given initial unit vector $v_0$:
$$
\forall v_0 \in \mathbb{R}^d,\left|v_0\right|=1 \quad \mathbb{P}\left\{\left|\log \left\|T_n v_0\right\|-L_n\right|>\varepsilon n\right\}<e^{-\delta n} .
$$

Finally, the expectations $L_n$ satisfy a lower bound
$$
L_n \geq n h,
$$
where the constant $h>0$ can be chosen uniformly for all possible sequences $\left\{\mu_n\right\} \in$ $\mathcal{K}^{\mathbb{N}}$.
\end{theorem}

As a corollary, we also have a parametric version of \autoref{noniidFurst}:

\begin{corollary}[Parametric, Non-Stationary Furstenberg Theorem, Theorem 2.2 in \cite{Gorodetski:2024aa}]\label{paraFurst}
Under the assumptions of \autoref{noniidFurst}, for any $\varepsilon>0$ there exists $\delta>0$ such that for all sufficiently large $n \in \mathbb{N}$ and all $E \in J$ we have
$$
\mathbb{P}\left\{\left|\log \left\|T_{n, E, \omega}\right\|-L_{n, E}\right|>\varepsilon n\right\}<e^{-\delta n},
$$
where $\mathbb{P}=\mu_1^E \times \mu_2^E \times \ldots \times \mu_n^E$. 
Moreover, the same estimate holds for the lengths of random images of any given initial unit vector $v_0$:
$$
\forall v_0 \in \mathbb{R}^2,\left|v_0\right|=1 \quad \mathbb{P}\left\{\left|\log \left\|T_{n, E, \omega} v_0\right\|-L_{n, E}\right|>\varepsilon n\right\}<e^{-\delta n} .
$$
\end{corollary}

To tie together large deviations estimates on transfer matrices to Green's functions, we require a new object to act as a bridge. Define the \textit{characteristic polynomial} of $H_{[a,b], \omega}$ as
\[P_{[a,b],E, \omega} := \det(H_{[a,b], \omega} - E)\word{for} a \leq b.\]
By convention, if $a > b$ set $P_{[a,b],E, \omega} = 1$. We view $P_{[a,b],E, \omega}$ as a $(b-a+1)$-degree polynomial in the variable $E$. A useful identity (see \cite{Damanik:2022aa} Proposition 2.2.9) that allows us to connect characteristic polynomials and Green's functions is
\begin{equation}\label{G_to_P_eq}
\abs{G_{[a,b], E, \omega}(x, y)} = \frac{\left|P_{[a, x-1], E, \omega} P_{[y+1, b], E, \omega}\right|}{\left|P_{[a, b], E, \omega}\right|} \word{for} x \leq y.
\end{equation}
Thus we will be controlling the behavior of the polynomials $P_{[a,b],E, \omega}$ to control growth of the generalized eigenfunctions. In turn, $P_{[a,b],E, \omega}$ and $T_{[a,b], E, \omega}$ are connected by \cite{Damanik:2022aa} Proposition 2.2.5:
\begin{equation}\label{polys_are_entries}
T_{[a,b], E, \omega} = \begin{bmatrix}
P_{[a,b], E, \omega} & - P_{[a+1, b], E, \omega}\\
P_{[a, b-1], E, \omega} & - P_{[a+1, b-1], E, \omega}\\
\end{bmatrix}.
\end{equation}
Our machinery requires a large deviations estimate on the polynomials $P_{[a,b],E, \omega}$. It is reasonable to expect such a result holds since the entries of our transfer matrices $T_{[a,b], E, \omega}$ are precisely these polynomials. This can be seen as a non-stationary extension of Theorem 2 in \cite{tsay99, Bucaj:2019aa}.

To utilize \autoref{paraFurst} to achieve such an estimate, we also require a result from Gorodetski, Kleptsyn, and Monakov about the H\"older-regularity of products of our distributions. We state the theorem as applicable in our context, but it holds for general random dynamical systems.

\begin{theorem}[Theorem 2.8 in \cite{Gorodetski:2022aa}]\label{Holder_reg}
Under the assumptions of \autoref{noniidFurst},
there exist $\alpha>0, C, \kappa<1$ such that for any initial measure $\nu_0$, any $n$, and any distributions $\mu_1, \ldots, \mu_n \in \mathcal{K}$ the $n$-th image of $\nu_0$ satisfies $(\alpha, C)$-Hölder property on the scales up to $\kappa^n$:
$$
\forall x \in \R^2 \quad \forall r>\kappa^n \quad\left(\mu_n \times \cdots \times \mu_1 \times \nu_0\right)\left(B_r(x)\right)<C r^\alpha .
$$
\end{theorem}

Now, we prove a large deviations estimate for the entries of our transfer matrices $T_{[a,b], E, \omega}$. 

\begin{theorem}\label{LDT_for_P}
Under the assumptions of \autoref{noniidFurst}, for any $\varepsilon>0$ there exists $\delta>0$ such that for all sufficiently large $b-a$, all $E \in J$, we have
$$
\mathbb{P}\left\{\left|\log \left|P_{[a,b],E, \omega} \right|-L_{[a,b], E}\right|>\varepsilon (b-a + 1)\right\}<e^{-\delta (b - a + 1)}.
$$
\end{theorem}

\begin{proof}
We begin by showing
\begin{equation} \label{LDE_for_n}
\mathbb{P}\left\{\left|\log \left|\langle T_{n, E, \omega}e_1, e_1\rangle \right|-L_{n, E}\right|>\varepsilon n\right\}<\tilde Ce^{-\delta n}
\end{equation}
for some constant $\tilde C$, some $\delta >0$, and sufficiently large $n$, where $e_1 = [1, 0]^T$. First, since $\left|\langle T_{n, E, \omega}e_1, e_1\rangle \right| \leq \norm{T_{n, E, \omega} e_1}^{1/2}$:
\begin{align*}
	\P\cparen{\log \abs{\langle T_{n, E, \omega}e_1, e_1\rangle}-L_{n, E} >\varepsilon n} & \leq\P\cparen{\log \paren{\norm{T_{n, E, \omega} e_1}^{1/2}} -L_{n, E} >\varepsilon n}\\
	& \leq\P\cparen{\log \norm{T_{n, E, \omega} e_1} -L_{n, E} >\varepsilon n}\\
	& < e^{-\delta n}
\end{align*}
by \autoref{paraFurst}.

For the other bound, consider the set inclusion
\[\{\left|\langle T_{n, E, \omega}e_1, e_1\rangle \right| \leq e^{L_{n, E} - \epsilon n}\} \subseteq \{\norm{T_{n, E, \omega}e_1} \leq e^{L_{n, E} - \frac{\epsilon}{2} n}\} \cup \cparen{\frac{\left|\langle T_{n, E, \omega}e_1, e_1\rangle \right|}{\norm{T_{n, E, \omega}e_1}} \leq e^{-\frac{\epsilon}{2}n}}\]
since if $\left|\langle T_{n, E, \omega}e_1, e_1\rangle \right| = \norm{T_{n, E, \omega}e_1}\cdot \frac{\left|\langle T_{n, E, \omega}e_1, e_1\rangle \right|}{\norm{T_{n, E, \omega}e_1}} \leq e^{L_{n, E} - \epsilon n}$ and $\norm{T_{n, E, \omega}e_1} > e^{L_{n, E} - \frac{\epsilon}{2} n}$ then 
\[\frac{\left|\langle T_{n, E, \omega}e_1, e_1\rangle \right|}{\norm{T_{n, E, \omega}e_1}} \leq e^{L_{n, E} - \epsilon n}\cdot e^{-L_{n, E} + \frac{\epsilon}{2} n} = e^{-\frac{\epsilon}{2} n}.\]

Therefore we have
\begin{align*}
	& \P \cparen{\log \left|\langle T_{n, E, \omega}e_1, e_1\rangle \right|-L_{n, E} \leq -\varepsilon n}\\
	& = \P\left\{ \left|\langle T_{n, E, \omega}e_1, e_1\rangle \right| \leq e^{L_{n, E} - \varepsilon n }\right\}\\
	& \leq \P\{\norm{T_{n, E, \omega}e_1} \leq e^{L_{n, E} - \frac{\epsilon}{2} n}\} + \P\cparen{\frac{\left|\langle T_{n, E, \omega}e_1, e_1\rangle \right|}{\norm{T_{n, E, \omega}e_1}} \leq e^{-\frac{\epsilon}{2}n}}.
\end{align*}
The first term is bounded by $e^{-\delta_0 n}$ for some $\delta_0 > 0$ associated to $\frac{\epsilon}{2}$ by \autoref{paraFurst}. To bound the second term, we invoke \autoref{Holder_reg}. Letting $e_2 = [0, 1]^T$:
\begin{align*}
	& \cparen{\frac{\left|\langle T_{n, E, \omega}e_1, e_1\rangle \right|}{\norm{T_{n, E, \omega}e_1}} \leq e^{-\frac{\epsilon}{2}n}} \\
	& = \cparen{\frac{\left|\langle T_{n, E, \omega}e_1, e_1\rangle \right|^2}{\norm{T_{n, E, \omega}e_1}^2} \leq e^{-\epsilon n}} \\
	& = \cparen{\frac{\norm{T_{n, E, \omega}e_1}^2 - \left|\langle T_{n, E, \omega}e_1, e_2\rangle \right|^2}{\norm{T_{n, E, \omega}e_1}^2} \leq e^{-\epsilon n}} \\
	& = \cparen{\frac{\left|\langle T_{n, E, \omega}e_1, e_2\rangle \right|^2}{\norm{T_{n, E, \omega}e_1}^2} \geq \paren{1- e^{-\epsilon n}}} \\
	& = \cparen{\paren{1 - e^{-\epsilon n}}^{1/2} \leq \frac{\abs{\langle T_{n, E, \omega} e_1, e_2 \rangle}}{\norm{T_{n, E, \omega}e_1}}}.
\end{align*}
Therefore, on this event, we have
\[\paren{1 - e^{-\epsilon n}}^{1/2} \leq \frac{\abs{\langle T_{n, E, \omega} e_1, e_2 \rangle}}{\norm{T_{n, E, \omega}e_1}} \leq 1 \word{and} 0 \leq \frac{\abs{\langle T_{n, E, \omega} e_1, e_1 \rangle}}{\norm{T_{n, E, \omega}}} \leq e^{-\frac{\epsilon}{2}n}.\]
Hence
\begin{align*}
	\norm{\frac{T_{n, E, \omega} e_1}{\norm{T_{n, E, \omega}e_1}} - e_2}^2 & = \paren{\frac{\abs{\langle T_{n, E, \omega} e_1, e_1 \rangle}}{\norm{T_{n, E, \omega}e_1}}}^2 + \paren{\frac{\abs{\langle T_{n, E, \omega} e_1, e_2 \rangle}}{\norm{T_{n, E, \omega}e_1}} - 1}^2\\
	& \leq e^{-\epsilon n} + ((1 - e^{-\epsilon n})^{1/2} - 1)^2\\
	& \leq e^{-\epsilon n} + (1 - e^{-\frac{\epsilon}{2} n} - 1)^2\\
	& = 2 e^{-\epsilon n}.
\end{align*}
Hence, when $\frac{\left|\langle T_{n, E, \omega}e_1, e_1\rangle \right|}{\norm{T_{n, E, \omega}e_1}} \leq e^{-\frac{\epsilon}{2}n}$, we have
\[\frac{T_{n, E, \omega} e_1}{\norm{T_{n, E, \omega}e_1}} \in B_{\sqrt{2}e^{-\epsilon n/2}}\paren{e_2}.\]
Hence
\begin{align*}
	\P\cparen{\frac{\left|\langle T_{n, E, \omega}e_1, e_1\rangle \right|}{\norm{T_{n, E, \omega}e_1}} \leq e^{-\frac{\epsilon}{2}n}} & \leq \P\cparen{\frac{T_{n, E, \omega} e_1}{\norm{T_{n, E, \omega}e_1}} \in B_{\sqrt{2}e^{-\epsilon n/2}}\paren{e_2}}\\
	& = \paren{\mu_n \times \cdots \times \mu_1 \times \delta_{e_1}}\cparen{\frac{T_{n, E, \omega} v}{\norm{T_{n, E, \omega}v}} \in B_{\sqrt{2}e^{-\epsilon n/2}}\paren{e_2}} 
\end{align*}
where $\mu_k$ is the distribution of $\frac{A_{k, E, \omega} (\cdot)}{\norm{A_{k, E, \omega} (\cdot)}} $.

Applying \autoref{Holder_reg}, set $\tilde{\kappa} = \max\cparen{\kappa, e^{-\epsilon/2}} < 1$ and $r = \sqrt{2}\tilde{\kappa}^n$. Then, with this choice of $r$:
\[(\mu_n \times \cdots \times \mu_1 \times \delta_{e_1})\paren{ B_{\sqrt{2}e^{-\epsilon n/2}}\paren{e_2}} \leq C \cdot r^{\alpha} \]
for some $\alpha > 0$. Hence 
\[\P\cparen{\frac{\left|\langle T_{n, E, \omega}e_1, e_1\rangle \right|}{\norm{T_{n, E, \omega}e_1}} \leq e^{-\frac{\epsilon}{2}n}} \leq C (\sqrt{2}^\alpha) \tilde{\kappa}^{n \alpha} = C (\sqrt{2}^\alpha) e^{n \alpha \log(\tilde{\kappa})} \]

Setting $\delta = \min\{\delta_0, \alpha \log(\tilde{\kappa})\}$ demonstrates \eqref{LDE_for_n}.

For the result, recall \eqref{polys_are_entries} above. Hence $P_{[1,n], E, \omega} = \langle T_{[1,n], E, \omega}e_1, e_1\rangle$ and the result holds for intervals of the form $[1,n]$. To get the result for general $[a,b]$, we reindex to get the desired bound.
\end{proof}

From this we yield a technical corollary that will be useful when proving the main result:

\begin{corollary}\label{max_corr}
For almost every $\omega$, for every $\epsilon >0$ and all $E \in J$ there is an $N(\omega, \epsilon, E) = N \in \N$ sufficiently large such for all $n \geq N$ we have
\[\max\cparen{\log\abs{P_{[n+1,2n], E, \omega}} - L_{[n+1,2n], E, \omega}, \log\abs{P_{[2n+2,3n+1], E, \omega}} - L_{[2n+2,3n+1], E, \omega}} < \epsilon n.\]
\end{corollary}

\begin{proof}
Focusing on the first term, by \autoref{LDT_for_P}, we know 
\[\mathbb{P}\left\{\left|\log \left|P_{[n+1,2n],E, \omega} \right|-L_{[n+1,2n], E, \omega}\right|>\varepsilon n\right\}<e^{-\delta n}\]
for some $\delta >0$. By Borel-Cantelli, 
\[\P\cparen{\left|\log \left|P_{[n+1,2n],E, \omega} \right|-L_{[n+1,2n], E, \omega}\right|>\varepsilon n \text{ infinitely often}} = 0.\]
Hence, for some $N_1 \in \N$ sufficiently large, $n \geq N_1$ implies
\[\left|\log \left|P_{[n+1,2n],E, \omega} \right|-L_{[n+1,2n], E, \omega}\right|\leq \varepsilon n \]
which of course implies
\[\log \left|P_{[n+1,2n],E, \omega} \right|-L_{[n+1,2n], E, \omega} \leq \varepsilon n .\]
The proof for the other term is similar and taking $N\geq N_1, N_2$ finishes the proof.
\end{proof}


\section{Properties of the Growth Functions}\label{Properties}

Now that we have a more concrete understanding of how $\log\norm{T_n}$ behaves with respect to its expectation, we can use this to establish that the family $\left\{\frac{1}{n} L_{n, E}\right\}_{n \in \mathbb{N}}$ is equicontinuous in the parameter $E \in J$ (\autoref{prelim_equicont}). One can view this as the non-stationary analog of continuity of the Lyapunov exponent. This fact will allow us to control the change of all our growth functions over a small interval which is crucial for \autoref{close_to_eigenvalues} where we analyze the structure of the sets where large deviations occur. This section follows closely the proof of Lemma 2.3 in \cite{Gorodetski:2024aa}.

We introduce some notation that will be useful in the following proofs. For any fixed $k \in \mathbb{N}$ we can decompose the product $T_{n, E, \omega}$ into products of groups of $k$ transfer matrices:
$$
T_{n, E, \omega}=B_m(E) \ldots B_1(E), \quad n = m \cdot k
$$
where
$$
B_j(E):=\left(A_{k j, E, \omega} \ldots A_{k(j-1)+1, E, \omega}\right), \quad j=1, \ldots, m .
$$
Now, take any unit vector $v_0$ and define
$$
\xi_{j, E}:=\log \left\|B_j(E)\right\|, \quad S_{j, E}:=\log \left|T_{k j, E, \omega}\left(v_0\right)\right|, \quad R_{j, E}=S_{j, E}+\xi_{j+1, E}-S_{j+1, E} .
$$
In this notation, we have

\begin{lemma}[Proposition 2.5 in \cite{Gorodetski:2024aa}]\label{equicont_lemma}
For any $\varepsilon^*>0$ there exists $k_1$, such that for any $k>k_1$ for some $\delta^*>0$ one has for all $n=k m$ and for any $E \in J$
$$
\mathbb{P}\left\{R_{0, E}+R_{1, E}+\cdots+R_{m-1, E}>n \varepsilon^*\right\}<e^{-\delta^* m}.
$$
\end{lemma}

We also require

\begin{lemma}[Part of Lemma 3.5 in \cite{Gorodetski:2023aa}]\label{uniform_bound_log_moments}
There exists $B$ such that for any $\mu \in \mathcal{K}$
\[\int\log\norm{A}d \mu(A)< B, \qquad \int\log^2\norm{A}d \mu(A)< B^2.\]
\end{lemma}

\begin{proof} First, note that $\frac{\log^2(x)}{x^\gamma}$ is uniformly bounded on $[1, \infty)$. Therefore $\frac{\log^2(\norm{A})}{\norm{A}^\gamma}$ is also uniformly bounded for any transfer matrix $A \in \mathrm{SL}(2, \R)$ as well. Hence the inequality follows from the uniform bound on the $\gamma$-moment.

Second, since $\mu$ is a probability measure:
\[\int\log\norm{A}d \mu(A) \leq \paren{\int\log^2\norm{A}d \mu(A)}^{1/2} < B\]
\end{proof}

\begin{lemma}\label{largeMlemma}
Fix $k \in \N$. Set
\[
\Omega_{M, j}=\left\{\omega \in \Omega: \exists i \in\left\{k(j-1)+1, \ldots, kj\right\} \text { such that }\left\|A_{i, E, \omega}\right\|>M\right\}
.\]
For all $\varepsilon>0$ and all $E \in J$ there exists $M = M(\epsilon) \gg 1$ such that
\[
\mathbb{P}\left(\Omega_{M, j}\right)<\frac{\varepsilon^2}{25\cdot k^2 \cdot B^2}
\]
where $M$ may be chosen uniformly with respect to $j$.
\end{lemma}

\begin{proof} Notice that
\begin{align*}
\mathbb{P}\left(\Omega_{M,j}\right) & \leq \sum_{i=k(j -1)+1}^{k j} \mathbb{P}\left\{\left\|A_{i, E, \omega}\right\|>M\right\} \\
& =\sum_{i=k(j-1)+1}^{k j} \mathbb{P}\left\{\left\|A_{i, E, \omega}\right\|^\gamma>M^\gamma\right\} \\
& \leq \sum_{i=k(j-1)+1}^{k j} \mathbb{E}\left[\left\|A_{i, E, \omega}\right\|^\gamma\right] \cdot M^{-\gamma} \\
& \leq k \cdot C \cdot M^{-\gamma} \\
& <\frac{\varepsilon^2}{2 5 \cdot k^2 \cdot B^2}
\end{align*}
where the last line follows by choosing $M$ appropriately large, since $\gamma >0$.
\end{proof}

In proving equicontinuity of $\left\{\frac{1}{n} L_n(E)\right\}_{n \in \mathbb{N}}$, we cut $T_{n, E, \omega}$ into $k$-sized blocks as above. The next lemma argues that, in the case when the potentials are bounded, we have equicontinuity of the blocks. 

\begin{lemma}\label{bounded_means_equicont}
Let $M, k \gg 1$ be given. For any $\varepsilon>0$ there exists $\delta>0$ such that for any $V_1, \ldots, V_k \in[-M, M]$, if $E_1, E_2 \in J$ and $\left|E_1-E_2\right|<\delta$, then
\[
\left\|A_{V_k, E_1} A_{V_{k-1}, E_1} \cdots A_{V_1, E_1}-A_{V_k, E_2} \cdots A_{V_1, E_2}\right\| \leq \varepsilon
\]
where $A_{V_i, E}$ denotes the transition matrix with potential $V_i$ and energy parameter $E$.
\end{lemma}

\begin{proof}
If $\abs{V_i} \leqslant M$ for all $i$, then $\left\|A_{V_k, E} \ldots A_{V_i, E}\right\|$ is also bounded in derivative since entries of $A_{V_k, E} \cdots A_{V_1, E}$ are polynomials of bounded degree with uniformly bounded coefficients on a compact interval $J$. Furthermore, the uniformity of this bound ensures equicontinuity for any choice of potentials $V_1, \ldots, V_k$.
\end{proof}

These lemmas will help us in establishing the following result:

\begin{theorem} \label{prelim_equicont} 
The sequence of functions $\left\{\frac{1}{n} L_{n, E}\right\}_{n \in \mathbb{N}}$ is equicontinuous (in the parameter $E \in J$).
\end{theorem}

\begin{proof}
Fix $k \in \mathbb{N}$. By the above definitions and iteratively applying $R_{j, E}=S_{j, E}+\xi_{j+1, E}-S_{j+1, E}$, we have
\begin{equation}\label{eq:equicont1}
\log \left|T_{k m, E, \omega} v_0\right|=S_{m, E}=\left(\xi_{1, E}+\cdots+\xi_{m, E}\right)-\left(R_{0, E}+R_{1, E}+\cdots+R_{m-1, E}\right)
\end{equation}
and hence, 
\begin{align*}
\log \left\|T_{n, E, \omega}\right\| & \geq \log \left|T_{n, E, \omega} v_0\right| \\
& =\left(\xi_{1, E}+\cdots+\xi_{m, E}\right)-\left(R_{0, E}+R_{1, E}+\cdots+R_{m-1, E}\right)\numberthis\label{eq:equicont2}.
\end{align*}
Fix $\varepsilon>0$ and take $\varepsilon^*:=\frac{\varepsilon}{5}$. By \autoref{equicont_lemma}, for any sufficiently large $k$ there exists $\delta^*>0$ such that for any $E \in J$
\begin{equation}\label{eq:equicont3}
R_{0, E}+R_{1, E}+\cdots+R_{m-1, E} < \frac{\varepsilon}{5} n
\end{equation}
with probability at least $1-e^{-\delta^* m}$. Once \eqref{eq:equicont3} holds, we have from \eqref{eq:equicont2}
\[\left(\xi_{1, E}+\cdots+\xi_{m, E}\right) - \log \left\|T_{n, E, \omega}\right\| \leq R_{0, E}+R_{1, E}+\cdots+R_{m-1, E} < \frac{\varepsilon}{5} n\]
and, by the sub-multiplicativity of the matrix norm, 
\[\log \left\|T_{n, E, \omega}\right\| - \left(\xi_{1, E}+\cdots+\xi_{m, E}\right) \leq 0 < \frac{\epsilon}{5} n\]
Therefore, putting these together,
$$
\left|\log \left\|T_{n, E, \omega}\right\|-\left(\xi_{1, E}+\cdots+\xi_{m, E}\right)\right|<\frac{\varepsilon}{5} n .
$$
Thus, for any two parameter values $E_1, E_2 \in J$ both inequalities
\begin{equation}\label{eq:equicont4}
\left|\log \left\|T_{n, E_i, \omega}\right\|-\left(\xi_{1, E_i}+\cdots+\xi_{m, E_i}\right)\right|<\frac{\varepsilon}{5} n, \quad i=1,2
\end{equation}
hold with the probability at least $1-2 e^{-\delta^* m}$.

Now, we wish to bound $\mathbb{E}\bracket{\abs{\xi_{j, E_1}-\xi_{j, E_2} }}$. To that end, notice
\[
\mathbb{E}\bracket{\abs{\xi_{j, E_1}-\xi_{j, E_2} }}=\mathbb{E}\bracket{\mathbb{1}_{\Omega_{M, j}} \cdot \abs{\xi_{j, E_1}-\xi_{j, E_2}}}+\mathbb{E}\bracket{\mathbb{1}_{\Omega_{M, j}^C} \cdot \abs{\xi_{j, E_1}-\xi_{j, E_2} }}
\]
where $\Omega_{M, j}$ is as in \autoref{largeMlemma}. By \autoref{bounded_means_equicont}, the second summand may be bounded by $\frac{\epsilon}{5}$ provided $|E_1 - E_2| < \delta$. Focusing on the first summand, note
\begin{align*}
	& \E\bracket{\mathbb{1}_{\Omega_{M,j}} \cdot \abs{\xi_{j, E_1} - \xi_{j, E_2}}}\\
	& = \E\bracket{\mathbb{1}_{\Omega_{M,j}} \cdot \abs{\log\norm{A_{kj,E_1,\omega}\cdots A_{k(j-1)+1,E_1,\omega}} - \log\norm{A_{kj,E_2,\omega}\cdots A_{k(j-1)+1,E_2,\omega}}}}\\
	& \leq \E\bracket{\mathbb{1}_{\Omega_{M,j}} \cdot \sum_{i=k(j-1)+1}^{kj} \log\norm{A_{i,E_1,\omega}}} + \E\bracket{\mathbb{1}_{\Omega_{M,j}} \cdot \sum_{i=k(j-1)+1}^{kj} \log\norm{A_{i,E_2,\omega}}}\\
	& = \sum_{i=k(j-1)+1}^{kj} \E\bracket{\mathbb{1}_{\Omega_{M,j}} \cdot \log\norm{A_{i,E_1,\omega}}} + \E\bracket{\mathbb{1}_{\Omega_{M,j}} \cdot \log\norm{A_{i,E_2,\omega}}}\\
	& \leq \sum_{i=k(j-1)+1}^{kj} \sqrt{\P\paren{\Omega_{M,j}}}\cdot \sqrt{\E\bracket{ \log^2\norm{A_{i,E_1,\omega}}}} + \sqrt{\P\paren{\Omega_{M,j}}}\cdot \sqrt{\E\bracket{ \log^2\norm{A_{i,E_2,\omega}}}}\\
	& < \sum_{i=k(j-1)+1}^{kj} \frac{\epsilon}{5\cdot k \cdot B}\cdot B + \frac{\epsilon}{5\cdot k \cdot B}\cdot B\\
	& = \frac{2}{5} \epsilon.
\end{align*}
Thus $\mathbb{E}\bracket{\abs{\xi_{j, E_1}-\xi_{j, E_2} }}<\frac{3}{5} \varepsilon$ provided that $|E_1 - E_2| < \delta$ where $\delta >0$ may be chosen uniformly with respect to $j \in \N$. Therefore, with $|E_1 - E_2| < \delta$, we have
\[\sum_{j=1}^m \mathbb{E}\bracket{\abs{\xi_{j, E_1}-\xi_{j, E_2} }} \leq \frac{3}{5}\cdot m \cdot \epsilon .\]
Combining this with \eqref{eq:equicont4} we have that 
\begin{equation}\label{eq:equicont5}
\E\bracket{\abs{\frac{1}{n} \log \norm{T_{n, E_1, \omega}} - \frac{1}{n} \log \norm{T_{n, E_2, \omega}}}} \leq \epsilon
\end{equation}
with probability at least $1 - 2e^{-\delta^* m}$. 

We are now ready to consider the expectations $L_{n, E_i}=\mathbb{E} \log \left\|T_{n, E_i, \omega}\right\|$. We have
\begin{align*}
	\frac{1}{n}\abs{L_{n, E_1} - L_{n, E_2}} & \leq  \E\bracket{\abs{\frac{1}{n} \log \norm{T_{n, E_1, \omega}} - \frac{1}{n} \log \norm{T_{n, E_2, \omega}}}}\\
	& = \E\bracket{\mathbb{1}_{\text{\eqref{eq:equicont5} holds}}\abs{\frac{1}{n} \log \norm{T_{n, E_1, \omega}} - \frac{1}{n} \log \norm{T_{n, E_2, \omega}}}}\\
	& \quad + \E\bracket{\mathbb{1}_{\text{\eqref{eq:equicont5} does not hold}}\abs{\frac{1}{n} \log \norm{T_{n, E_1, \omega}} - \frac{1}{n} \log \norm{T_{n, E_2, \omega}}}}.
\end{align*}
The first summand where \eqref{eq:equicont5} holds is evidently bounded by $\epsilon$. For the second summand, note
\begin{align*}
	& \E\bracket{\mathbb{1}_{\text{\eqref{eq:equicont5} does not hold}}\abs{\frac{1}{n} \log \norm{T_{n, E_1, \omega}} - \frac{1}{n} \log \norm{T_{n, E_2, \omega}}}}\\
	& \leq \E\bracket{\mathbb{1}_{\text{\eqref{eq:equicont5} does not hold}}\abs{\frac{1}{n} \log \norm{T_{n, E_1, \omega}}}}\\
	& \quad  + \E\bracket{\mathbb{1}_{\text{\eqref{eq:equicont5} does not hold}}\abs{\frac{1}{n} \log \norm{T_{n, E_2, \omega}}}}\\
	& \leq \frac{1}{n} \sum_{i = 1}^n \sqrt{\P\paren{\text{\eqref{eq:equicont5} does not hold}}} \sqrt{\E[\log^2\norm{A_{i, E_1, \omega}}]} \\
	& \quad + \frac{1}{n} \sum_{i = 1}^n\sqrt{\P\paren{\text{\eqref{eq:equicont5} does not hold}}} \sqrt{\E[\log^2\norm{A_{i, E_2, \omega}}]}\\
	& \leq \frac{2}{n} \sum_{i = 1}^n 2e^{-\delta^* m} \cdot B\\
	& = 4e^{-\delta^* m} \cdot B
\end{align*}
which tends to 0 as $m \to \infty$. With an easy handling of $n$ not divisible by $k$ and of a
finite number of $n$ that are too small we obtain the desired equicontinuity.
\end{proof}

\begin{corollary}
\label{growth_function_equicontinuty}
The sequence of functions $\left\{\frac{1}{b-a + 1} L_{[a,b], E}\right\}$ is equicontinuous (in the parameter $E \in J$).
\end{corollary}

One consequence of equicontinuity that is useful to leverage is that we can now control how growth functions over adjacent windows behave. For example, if we consider $L_{[1,n], E}$, we can estimate how far this from $L_{[1,c], E} + L_{[c+1,n], E}$ for some $c$ between 1 and n. Equicontinuity allows us to do this in away that is uniform over $E \in J$. 

Before we can prove such a result, we need to control what is happening to ``small'' windows like $[1,c]$ in the discussion above.

\begin{lemma} \label{window_prelim}
For any $\varepsilon^{\prime}>0$ and any $E \in J$ there exists $\delta'>0$ such that for all sufficiently large $n \in \mathbb{N}$ the following holds: with probability at least $1-\exp \left(-\delta' n\right)$ for all $a, b$ with $0 \leq a<b\leq n$ one has
\[
\abs{\log \left\|T_{\left[a, b\right], E, \omega}\right\| - L_{\left[a, b\right], E}} \leq n\epsilon'.
\]
\end{lemma}

\begin{proof}
Fix $\epsilon >0$ and $E \in J$. The idea of the proof is to consider when $b-a + 1$ is ``big'' and when it is ``small'' relative to $n$ in two different cases. When $b-a+1$ is large, the proof proceeds in much the same way as the proof of Corollary 2.10 in \cite{Gorodetski:2024aa}. When $b-a+1$ is small, the argument in \cite{Gorodetski:2024aa} relies on unformly bounded potentials. To handle this case, we instead leverage the finite $\gamma$-moment.

First consider the ``$b-a+1$ is large'' case; suppose $b-a + 1 \geq \frac{\gamma n \epsilon}{2\paren{\log(C) + \gamma B}}$, where $B > 0$ is the linear upper bound on the growth functions as in \autoref{uniform_bound_log_moments} (so $L_{[a,b], E} \leq (b-a+1)\cdot B$ for all $a \leq b$). In this case, if we have
\[
\abs{\log \left\|T_{\left[a, b\right], E, \omega}\right\|-L_{\left[a, b\right], E}} > n \epsilon
\]
then $b-a+1 \leq n$ implies
\[
\abs{\log \left\|T_{\left[a, b\right], E, \omega}\right\|-L_{\left[a, b\right], E}} > (b-a+1) \epsilon.
\]
The probability of the latter event happening is at most $e^{-\delta(b-a+1)} < e^{-\delta \frac{\gamma \epsilon}{2\paren{\log(C) + \gamma B}} n}$ by \autoref{paraFurst}. Allowing $a$ and $b$ to range through $[0, n]$, there are less than $n^2$ such events. The probability that at least one of them occurs is thus bounded by $n^2e^{-\delta \frac{\gamma \epsilon}{2\paren{\log(C) + \gamma B}}n}$. So the probability of $\abs{\log \left\|T_{\left[a, b\right], E, \omega}\right\|-L_{\left[a, b\right], E}} > n \bar{\varepsilon}$ for any $a$ and $b$ in this case is bounded by $n^2e^{-\delta \frac{\gamma \epsilon}{2\paren{\log(C) + \gamma B}}n}$.

Now consider the ``$b-a+1$ is small'' case. Suppose $b-a+1 < \frac{\gamma n \epsilon}{2\paren{\log(C) + \gamma B}}$. To bound the probability of 
\[\abs{\log \left\|T_{\left[a, b\right], E, \omega}\right\|-L_{\left[a, b\right], E}} > n \epsilon,\]
we bound the probability of the super- and sub-deviations separately.

First we compute a bound on the probability of super-deviations:
\begin{align*}
	\P \cparen{\log\norm{T_{[a,b], E, \omega}} - L_{[a,b],E} > n\cdot \epsilon} & = \P \cparen{\log\norm{T_{[a,b], E, \omega}} > n\cdot \epsilon + L_{[a,b],E}}\\
	& = \P \cparen{\norm{T_{[a,b], E, \omega}}^\gamma > \exp\bracket{\gamma(n\cdot \epsilon + L_{[a,b],E})}}\\
	& \leq \E\bracket{\norm{T_{[a,b], E, \omega}}^\gamma} \exp\bracket{-\gamma(n\cdot \epsilon + L_{[a,b],E})} \numberthis \label{window_prelim_eq1}\\
	& \leq \E\bracket{\prod_{i=a}^b\norm{A_{i, E, \omega}}^\gamma} \exp\bracket{-\gamma(n\cdot \epsilon + L_{[a,b],E})} \\
	& = \prod_{i=a}^b\E\bracket{\norm{A_{i, E, \omega}}^\gamma} \exp\bracket{-\gamma(n\cdot \epsilon + L_{[a,b],E})} \numberthis \label{window_prelim_eq2}\\
	& \leq C^{b-a+1} \exp\bracket{-\gamma(n\cdot \epsilon + L_{[a,b],E})} \\
	& = \exp\bracket{-n\gamma \epsilon + (b-a+1)\log(C) - \gamma L_{[a,b],E}}
\end{align*}
where \eqref{window_prelim_eq1} is Markov's inequality and \eqref{window_prelim_eq2} is from the independence of the transfer matrices $A_{i, E \omega}$. We can bound the probability of sub-deviations in a similar fashion:
\begin{align*}
	&\P \cparen{-(\log\norm{T_{[a,b], E, \omega}} - L_{[a,b],E}) > n\cdot \epsilon} \\
	& = \P \cparen{-\log\norm{T_{[a,b], E, \omega}} > n\cdot \epsilon - L_{[a,b],E}}\\
	& = \P \cparen{\norm{T_{[a,b], E, \omega}}^{-\gamma} > \exp\bracket{\gamma(n\cdot \epsilon - L_{[a,b],E})}}\\
	& \leq \E\bracket{\norm{T_{[a,b], E, \omega}}^{-\gamma}} \exp\bracket{-\gamma(n\cdot \epsilon - L_{[a,b],E})} \numberthis \label{window_prelim_eq3}\\
	& \leq \E\bracket{\norm{T_{[a,b], E, \omega}}^{\gamma}} \exp\bracket{-\gamma(n\cdot \epsilon - L_{[a,b],E})} \numberthis \label{window_prelim_eq4}\\
	& \leq C^{b-a+1} \exp\bracket{-\gamma(n\cdot \epsilon - L_{[a,b],E})} \\
	& = \exp\bracket{-n\gamma \epsilon + (b-a+1)\log(C) + \gamma L_{[a,b],E}}\\
	& \leq \exp\bracket{-n\gamma \epsilon + (b-a+1)(\log(C) + \gamma B)}
\end{align*}
where \eqref{window_prelim_eq3} is Markov's inequality again and \eqref{window_prelim_eq4} holds since $\norm{T_{[a,b], E, \omega}} \geq 1$ for all $a\leq b$, hence $\norm{T_{[a,b], E, \omega}}^{-\gamma} \leq 1 \leq \norm{T_{[a,b], E, \omega}}^{\gamma}$.  

Since $L_{[a,b],E} \geq 0$ for all $a \leq b$, we know $(b-a+1)(\log(C) + \gamma B) \geq (b-a+1)\log(C) - \gamma L_{[a,b],E}$. So we have the same bound on the super-deviations as the sub-deviations:
\[\P \cparen{\log\norm{T_{[a,b], E, \omega}} - L_{[a,b],E} > n\cdot \epsilon} \leq \exp\bracket{-n\gamma \epsilon + (b-a+1)(\log(C) + \gamma B)}.\] 
Since we supposed $b-a+1 < \frac{\gamma n \epsilon}{2\paren{\log(C) + \gamma B}}$, we have
\begin{align*}
	& \P \cparen{\abs{\log\norm{T_{[a,b], E, \omega}} - L_{[a,b],E}} > n\cdot \epsilon}\\
	& \leq \exp\bracket{-n\gamma \epsilon + (b-a+1)\paren{\log(C) + \gamma B}}\\
	& \leq \exp\bracket{-n\gamma \epsilon + \frac{\gamma n \epsilon}{2\paren{\log(C) + \gamma B}}\paren{\log(C) + \gamma B}}\\
	& = \exp\bracket{-\frac{\epsilon\gamma}{2} n}
\end{align*}
 
Finally combining these two cases, choosing $\delta'$ sufficiently small so that $n^2e^{-\delta \frac{\gamma \epsilon}{2D} n} \leq e^{-\delta' n}$ and $e^{-\frac{\epsilon\gamma}{2} n} \leq e^{-\delta' n}$ gives the result.
\end{proof}

We are now able to address the case when one considers different ``windows'' for our growth functions:

\begin{corollary}\label{growth_funct_rel_2} 
For any $\epsilon>0$ there exists $N \in \mathbb{N}$ such that for any $n \geq N$, any $a, b, c \in \mathbb{N}$ with $0 \leq a<b<c \leq n$, and any $E \in J$ we have
$$
0 \leq L_{\left[a, b\right], E, \omega}+L_{\left[b, c\right], E, \omega}-L_{\left[a, c\right], E, \omega} \leq n \epsilon.
$$
\end{corollary}

The proof for this follows from \autoref{growth_function_equicontinuty} and \autoref{window_prelim} in the same way as Proposition 3.8 in \cite{Gorodetski:2024aa}. The idea is to leverage the fact that, for a unit vector $v$,
\[
T_{\left[a, c\right], E, \omega} v=T_{\left[b, c\right], E, \omega} T_{\left[a, b\right], E, \omega} v,
\]
and hence
\[
\log \left\|T_{\left[a, c\right], E, \omega} v\right\|=\log \left\|T_{\left[b, c\right], E, \omega}\left(\frac{T_{\left[a, b\right], E, \omega} v}{\left\|T_{\left[a, b\right], E, \omega} v\right\|}\right)\right\|+\log \left\|T_{\left[a, b\right], E, \omega} v\right\|.
\]
Therefore with large probability we can argue that $L_{\left[a, b\right], E, \omega}+L_{\left[b, c\right], E, \omega}-L_{\left[a, c\right], E, \omega} \leq n \epsilon$. By chopping up $J$ into intervals of size no larger than $\delta$ and invoking equicontinuity, we can make this estimate uniform in $E \in J$.


\section{Analysis of the Large Deviation Sets}\label{Technical Lemmas}

At this point, we have a good understanding of how $\log\norm{T_n}$ behaves on sets of large probability. To complete the proof of \autoref{main_theorem_reform}, we also need to understand what happens in the sets where we do have large deviations. More concretely, we need to make a connection between sets of large deviations and points of singularity with respect to the Green's functions (\autoref{tech_cor_1}). We also need to understand how eigenvalues of the truncated operator behave within these large deviation sets (\autoref{close_to_eigenvalues} and \autoref{tech_lem_3}). The remaining lemmas will be useful when we break into three cases in the course of our proof of \autoref{main_theorem_reform} in Section \ref{Proof of Main Theorem}. Many of these lemmas either follow very closely or are identical to the lemmas in \cite{Rangamani:2019aa} with proofs modified as necessary to fit our context.

First, define the ``large deviation'' sets:
\[B_{[a, b], \varepsilon}^{+}=\left\{(E, \omega) : \log\abs{P_{[a, b], E, \omega}} - L_{[a, b], E} \geq (b - a + 1)\epsilon\right\}\] 
\[B_{[a, b], \varepsilon}^{-}=\left\{(E, \omega) : \log\abs{P_{[a, b], E, \omega}} - L_{[a, b], E} \leq -(b - a + 1)\epsilon\right\}\]
where we view $B^+$ as the ``super-deviations'' set and $B^-$ as the ``sub-deviations'' set. Denote the cross-sections of these sets as
\[B_{[a, b],\epsilon , \omega}^{ \pm}=\left\{E : (E, \omega) \in B_{[a, b], \varepsilon}^{ \pm}\right\}\]
\[B_{[a, b], \varepsilon, E}^{ \pm}=\left\{\omega : (E, \omega) \in B_{[a, b], \varepsilon}^{ \pm}\right\}\]
and lastly set $B_{[a, b], *}=B_{[a, b], *}^{+} \cup B_{[a, b], *}^{-}$. 

Notice that if we re-expressed, say, the sub-deviations set (with fixed $\omega$) as 
\[\cparen{E : \abs{P_{[a,b], E, \omega}} \leq e^{L_{[a, b], E} - (b-a+1)\epsilon}},\]
it is evident that the roots of our characteristic polynomial $P_{[a,b], E, \omega}$ are contained in these sets. Actually, since everything is continuous, an interval about the set is contained in the sub-deviation sets. We would like to claim that $B^{-}_{[a, b], \epsilon, \omega}$ comprises entirely of these intervals about roots of the characteristic polynomial. This is the result in \autoref{close_to_eigenvalues}, and this goal motivates the following reasoning.

Due to the equicontinuity of $\left\{\frac{1}{b-a + 1} L_{[a,b], E}\right\}$ proved in the previous section, without loss of generality we may take $J$ small enough to ensure that 
\[\sup_{a<b}\cparen{\max\cparen{\frac{1}{b-a + 1} L_{[a,b], E} : E \in J} - \min\cparen{\frac{1}{b-a + 1} L_{[a,b], E} : E \in J}} \leq \frac{\epsilon}{2}.\]
Proceed with this choice of $J$. Set 
\[\rho(a,b) = \frac{1}{2}\paren{\max\cparen{\frac{1}{b-a + 1} L_{[a,b], E} : E \in J} + \min\cparen{\frac{1}{b-a + 1} L_{[a,b], E} : E \in J}}\]
so that 
\[\sup_{a<b} \abs{\frac{1}{b-a+1} L_{[a,b], E} - \rho(a, b)} \leq \frac{\epsilon}{4}. \]
With this notation we establish the following fact:

\begin{lemma}\label{close_to_eigenvalues}
The set
\[B^{-}_{[n+1, 3n + 1], \epsilon, \omega} = \cparen{E : \abs{P_{[n+1,3n+1], E, \omega}} \leq e^{L_{[n+1, 3n+1], E} - (2n + 1)\epsilon}}\]
consists of at most $2n +1$ intervals about the roots of $P_{[n+1,3n+1], E, \omega}$.
\end{lemma}
 
\begin{proof}
Certainly such intervals are contained in $B^{-}_{[n+1, 3n + 1], \epsilon_0, \omega}$. To see that these are all the intervals, note that the level set
\[\{E : \abs{P_{[n+1,3n+1], E, \omega}} = e^{(2n+1)(\rho(n+1, 3n+1) - \frac{3}{4}\epsilon)}\}\]
consists of at most $2(2n+1)$ points since $P_{[n+1,3n+1], E, \omega}$ is degree $2n + 1$ and all the roots are simple (see \cite{Damanik:2022aa} Corollary 2.2.3). Therefore 
\begin{equation}\label{close_to_eigenvalues_eq1}
\{E : \abs{P_{[n+1,3n+1], E, \omega}} < e^{(2n+1)(\rho(n+1, 3n+1) - \frac{3}{4}\epsilon)}\}
\end{equation}
consists of at most $2n + 1$ open intervals between these points. Noticing $B^{-}_{[n+1, 3n + 1], \epsilon, \omega}$ is contained in the above set finishes the proof since simplicity of the roots of $P_{[n+1,3n+1], E, \omega}$ ensures there is at most one one interval in $B^{-}_{[n+1, 3n + 1], \epsilon, \omega}$ per interval contained in (\ref{close_to_eigenvalues_eq1}). Indeed, $L_{[n+1, 3n+1], E} - (2n + 1)\epsilon \leq (2n+1)\paren{\rho(n+1, 3n+1) + \frac{\epsilon}{4}} - (2n + 1)\epsilon = (2n+1)(\rho(n+1, 3n+1) - \frac{3}{4}\epsilon)$, so
\[\abs{P_{[n+1,3n+1], E, \omega}} \leq e^{L_{[n+1, 3n+1], E} - (2n + 1)\epsilon} \leq e^{(2n+1)(\rho(n+1, 3n+1) - \frac{3}{4}\epsilon)}.\]
\end{proof}

For the next lemma, we want to connect the behavior of the Green's functions to the large deviation sets above. Recall that $h$ is the constant provided by \autoref{noniidFurst} such that $L_n \geq n \cdot h$ for all $n$.

\begin{lemma} \label{tech_lem_1} 
Suppose $0<\epsilon < \frac{h}{5}$. For some $N \in \N$, $n \geq N$, if we have 
\[\log \abs{G_{[x - n, x + n], E, \omega}(x, x-n)} \geq -n(h - 5 \epsilon) + \epsilon\]
or 
\[\log \abs{G_{[x - n, x + n], E, \omega}(x, x + n)} \geq -n(h - 5 \epsilon) + \epsilon\]
then $(E, \omega) \in B^{-}_{[x-n, x+n], \epsilon} \cup B^{+}_{[x-n, x-1], \epsilon} \cup B^{+}_{[x+1, x+n], \epsilon}$.
\end{lemma}

One may notice that the windows (i.e., intervals) in the subscripts of the large deviation sets are not symmetric. The reason for this behavior is the structure of the entries of the Green's function as in \eqref{G_to_P_eq}; the windows derive from this identity.

\begin{proof}
We proceed by contrapositive; suppose 
\[(E, \omega) \in \paren{B^{-}_{[x-n, x+n], \epsilon}}^{C} \cap \paren{B^{+}_{[x-n, x-1], \epsilon}}^{C} \cap \paren{B^{+}_{[x+1, x+n], \epsilon}}^{C}\]
where, for clarity, 
\begin{align*}
	& \paren{B^{-}_{[x-n, x+n], \epsilon}}^{C} = \cparen{\omega : \log\abs{P_{[x-n, x+n], E, \omega}} - L_{[x-n, x+n], E} > -(2n-1)\epsilon},\\
	& \paren{B^{+}_{[x-n, x-1], \epsilon}}^{C} = \cparen{\omega : \log\abs{P_{[x-n, x-1], E, \omega}} - L_{[x-n, x-1], E} < n \epsilon},\\
	& \paren{B^{+}_{[x+1, x+n], \epsilon}}^{C} = \cparen{\omega : \log\abs{P_{[x+1, x+n], E, \omega}} - L_{[x+1, x+n], E} < n \epsilon}.
\end{align*}
Since $G_{[a,b], E, \omega}$ is symmetric, we have
\begin{align*}
	\log \abs{G_{[x - n, x + n], E, \omega}(x, x-n)} & = \log \abs{G_{[x - n, x + n], E, \omega}(x-n, x)}\\
	& = \log \abs{\frac{P_{[x-n, x-n-1], E, \omega} \cdot P_{[x+1, x + n], E, \omega}}{P_{[x - n, x + n], E, \omega}}} & \numberthis \label{tech_lem_1_eq1}\\
	& = \log \abs{\frac{P_{[x+1, x + n], E, \omega}}{P_{[x - n, x + n], E, \omega}}} & \numberthis \label{tech_lem_1_eq2}\\
	& = \log\abs{P_{[x+1, x + n], E, \omega}} - \log\abs{P_{[x-n, x + n], E, \omega}}\\
\end{align*}
where \eqref{tech_lem_1_eq1} is from \eqref{G_to_P_eq} and \eqref{tech_lem_1_eq2} follows since $x-n > x -n -1$. Applying our hypotheses:
\begin{align*}
	& \log\abs{P_{[x+1, x + n], E, \omega}} - \log\abs{P_{[x-n, x + n], E, \omega}} \\
	& < L_{[x+1, x+n], E} + n \epsilon - L_{[x - n, x + n], E} + (2n + 1) \epsilon\\
	& = (L_{[x+1, x+n], E} - L_{[x - n, x + n], E}) + (3n + 1) \epsilon\\
	& < 2n\epsilon - L_{[x-n, x + 1], E}+ (3n + 1) \epsilon & \numberthis \label{tech_lem_1_eq3}\\
	& = - L_{[x-n, x + 1], E} + (5n + 1)\epsilon
\end{align*}
where \eqref{tech_lem_1_eq3} follows from \autoref{growth_funct_rel_2}. Thus we have
\[\log \abs{G_{[x - n, x + n], E, \omega}(x, x-n)} < - L_{[x-n, x+1], E} + (5n + 1) \epsilon.\]
Similarly, we find
\[\log \abs{G_{[x - n, x + n], E, \omega}(x, x + n)} < - L_{[x-1, x+n], E} + (5n + 1)\epsilon.\]
Now, from \autoref{noniidFurst}, we know $L_{[a,b], E, \omega} \geq (b - a + 1) \cdot h$, where $h > 0$ is uniformly chosen. Hence
\[- L_{[x-n, x+1], E}, - L_{[x-1, x+n], E} \leq -(n+2)h \leq -n h\]
and so
\[\log \abs{G_{[x - n, x + n], E, \omega}(x, x-n)}, \log \abs{G_{[x - n, x + n], E, \omega}(x, x + n)} < -n(h - 5 \epsilon) + \epsilon\]
as desired.
\end{proof}

\begin{corollary}\label{tech_cor_1}
Let $\epsilon > 0$. For $n$ sufficiently large, if $2n+1$ is $(h - 6 \epsilon, n, E, \omega)$-singular, i.e., if
\[\abs{G_{[n+1, 3n+1], E, \omega}(2n+1, n+1)} \geq e^{-(h-6\epsilon) n}\]
or
\[\abs{G_{[n+1, 3n+1], E, \omega}(2n+1, 3n+1)} \geq e^{-(h-6\epsilon)n}\]
then $E \in B^{-}_{[n+1, 3n+1], \epsilon, \omega}$.
\end{corollary}

\begin{proof}
\autoref{max_corr} implies that $B^{+}_{[n+1, 2n], \epsilon}$ and $B^{+}_{[2n+2, 3n+1], \epsilon}$ are empty for sufficiently large $N$. Thus this corollary follows from \autoref{tech_lem_1}.
\end{proof}

Let $m$ denote Lebesgue measure on $\R$. The following lemma will allow us to control the size of the sets of large deviations with respect to Lebesgue measure which we mainly use to control how far apart eigenvalues of certain truncated operators can be in the proof of \autoref{main_theorem_reform}.

\begin{lemma}
\label{tech_lem_2}
Suppose $0<\epsilon$ and let $\delta$ be the corresponding large deviation parameter (from \autoref{LDT_for_P}), and $0<\eta<\delta$, then for almost every $\omega$, there is $N_1(\omega)$ such that for $n>N_1,$ $\max \left\{m\left(B_{[n+1,3 n+1], \epsilon, \omega}^{-}\right), m\left(B_{[-n, n], \epsilon, \omega}^{-}\right)\right\} \leq e^{-(\delta-\eta)(2 n+1)}$.
\end{lemma}

\begin{proof}
We have
\begin{align*}
\mathbb{E}\left(m\left(B_{[a, b], \epsilon, \omega}^{-}\right)\right) & =\int_{\mathbb{R}} \mathbb{P}\left(B_{[a, b], \epsilon, E}^{-}\right) d m(E) \\
& \leq m(J) e^{-\delta(b-a+1)}.
\end{align*}
The first equality is Fubini's theorem, and the second line follows by \autoref{LDT_for_P}.
Let
$$
F_n = \left\{\omega: m\left(B_{[n+1,3 n+1], \epsilon, \omega}^{-} \right) \geq e^{-\left(\delta-\eta\right)(2 n+1)}\right\},
$$
and
$$
G_n = \left\{\omega: m\left(B_{[-n,n], \epsilon, \omega}^{-} \right) \geq e^{-\left(\delta-\eta\right)(2 n+1)}\right\}.
$$
Note, by Markov's inequality:
\[\P(F_n) \leq \frac{\E\bracket{m\paren{B_{[n+1,3 n+1], \epsilon, \omega}^{-}}}}{e^{-\left(\delta-\eta\right)(2 n+1)}}\]
and hence, by the work above,
\[\P(F_n) \leq m(J)e^{-\eta(2n + 1)}.\]
Similarly, $\P(G_n) \leq m(J)e^{-\eta(2n + 1)}$. Thus,
$$
\P\left(F_n \cup G_n\right) \leq 2 m(J) e^{-\eta(2n + 1)}
$$
and the result follows by Borel-Cantelli.
\end{proof}

The next three lemmas are technical but will be useful for making estimates when we break the proof of \autoref{main_theorem_reform} into 3 cases.

\begin{lemma} 
\label{tech_lem_3}
Suppose $0<\varepsilon$, let $\delta$ be the corresponding large deviation parameter (from \autoref{LDT_for_P}) and $p>3 / \delta$. Let $E_j$ be the $j^\text{th}$ eigenvalue of $H_{[n+1,3n+1],\omega}$. For $n \in \mathbb{N}$, put 
\[C_n=\{\omega : \exists y \in [-n, n], |-n -y| \geq \ln \left(n^p\right)\text{, and }E_j \in B_{[-n, y], \epsilon, \omega}\text{ for some }1 \leq j \leq 2 n+1\},\]
and
\[D_n=\{\omega : \exists y \in [-n, n], |n -y| \geq \ln \left(n^p\right)\text{, and }E_j \in B_{[y, n], \epsilon, \omega}\text{ for some }1 \leq j \leq 2 n+1\}.\]
Then, $P\left[C_n \cup D_n \text{ infinitely often}\right]=0$.
\end{lemma}

\begin{proof}
Fix $n \in \mathbb{N}$ and $y$ with $|-n-y| \geq \ln \left(n^p\right)$, and $1 \leq j \leq 2 n+1$. Set
\[A_{y, j}=\left\{\omega: E_j \in B_{[-n, y], \epsilon, \omega}\right\} = B_{[-n, y], \epsilon, E_j}.\]
By \autoref{LDT_for_P} we have
\[\mathbb{P}\left(B_{[-n, y], \epsilon, E_j}\right) \leq e^{-\delta|-n-y|}.\]
Now for each $n$, if $Q_n^{\prime}=\left\{y \in[-n, n]:|-n-y| \geq \ln \left(n^p\right)\right\}$,
\[C_n=\bigcup_{\substack{y \in Q_n',\\ 1 \leq j \leq 2 n+1}} A_{y, j}.\]
By the above, we have $\mathbb{P}(C_n) \leq(2 n+1)^2 e^{-\delta \ln(n^p)} = (2n+1)^2 \cdot n^{-p\delta}$. Thus, $\mathbb{P}(C_n\text{ infinitely often})=0$ by Borel-Cantelli. The result follows by applying the same argument to $D_n$.
\end{proof}

\begin{lemma}
\label{tech_lem_4}
Suppose $p>0$ and $r>1$. Let 
\[J_n=\left\{\omega: \exists k \in[-n, n], \ \text{where }|-n-k| \leq \ln \left(n^p\right)\text{ or }|k-n| \leq \ln \left(n^p\right), \text{ and }\left|V_\omega(k)\right| \geq n^{r / \gamma} \right\}.\] 
Then, $\mathbb{P}\left(J_n \text{ infinitely often}\right)=0$.
\end{lemma}

\begin{proof}
Set $Q_n=\left\{k \in[-n, n] : |-n-k| \leq \ln \left(n^p\right)\right.$ or $\left.|n-k| \leq \ln \left(n^p\right)\right\}$, and $A_{k,n} = \{\omega:\left|V_\omega(k)\right| \geq n^{r / \gamma}\}$. Then,
$$
J_n=\bigcup_{k \in Q_n} A_{k,n}.
$$
By the Markov inequality:
\begin{align*}
	\P\paren{A_{k,n}} & = \P\cparen{\left|V_\omega(k)\right|^\gamma \geq n^{r}}\\
	& \leq \frac{\E[\left|V_\omega(k)\right|^\gamma]}{n^r}\\
	& < \frac{C}{n^r}
\end{align*}
where the last line follows from assumption \textbf{(A2)}. Thus, $\mathbb{P}\left(J_n\right) \leq 2C\left(\ln \left(n^p\right)+1\right) n^{-r}$. Since $p$ is fixed, this is certainly summable in $n$. By Borel-Cantelli, $\mathbb{P}\{J_n\text{ infinitely often}\}=0$.
\end{proof}

\begin{corollary} 
\label{tech_lem_5}
If $p>0$ and $r>1$, for almost every $\omega$, there is $N(\omega)$ such that for $n>N$ and any $k \in[-n, n]$ such that $|-n-k| \leq \ln \left(n^p\right)$ (respectively, $|n-k| \leq \ln \left(n^p\right)$),
\[\left|P_{[-n, k], E, \omega}\right| \leq n^{p(\ln(n^p) + 1)+ r/\gamma}\]
(respectively, $\left|P_{[k, n], E, \omega}\right| \leq n^{p(\ln(n^p) + 1)+ r/\gamma}$).
\end{corollary}

\begin{proof}
Note that the proof of \autoref{tech_lem_4} only utilized the finite $\gamma$-moment assumption for the potentials. Since $J$ is compact, $\max\cparen{\sup_{E \in J}\{V_\omega(j) - E\}, 1}$ also has a bound on the $\gamma$-moment, uniform in $j$. Hence, with appropriate assumptions as in \autoref{tech_lem_4} we have
\[\abs{\max\cparen{\sup_{E \in J}\{V_\omega(k) - E\}, 1}} < n^{r/\gamma}\]
for $n$ sufficiently large and $k$ such that $|n \pm k| \leq \ln(n^p)$. 

Now, using the combinatorial definition of determinant, we have
\begin{align*}
	\abs{P_{[-n,k], E, \omega}} & = \abs{\det\paren{H_{[-n,k], \omega} - E}}\\
	& = \abs{\sum_{\tau \in S_{n + k}} (-1)^{|\tau|} \prod_{i = 1}^{n + k} \paren{H_{[-n,k], \omega} - E}_{i, \tau(i)}}\\
	& \leq \sum_{\tau \in S_{n + k}} \prod_{i = 1}^{n + k} \abs{\paren{H_{[-n,k], \omega} - E}_{i, \tau(i)}}\\
	& \leq \sum_{\tau \in S_{n + k}} \prod_{i = 1}^{n + k} \abs{\max\cparen{\sup_{E \in J}\{V_\omega(i - (n+1)) - E\}, 1}}\\
	& \leq \abs{S_{n + k}}\cdot (n + k) \cdot \max_{j \in [-n, k]}\cparen{\sup_{E \in J}\{V_\omega(j) - E\}, 1}\\
	& \leq \ln(n^p)! \cdot (n + k) \cdot n^{r/\gamma}\\
	& \leq \ln(n^p)^{\ln(n^p) + 1}\cdot n^{r/\gamma}\\
	& \leq n^{p(\ln(n^p) + 1)}\cdot n^{r/\gamma}\\
	& = n^{p(\ln(n^p) + 1)+ r/\gamma}
\end{align*}
The proof for $\left|P_{[k, n], E, \omega}\right|$ is similar.
\end{proof}


\section{Proof of Spectral Localization}\label{Proof of Main Theorem}

We wish to remind the reader that, in order to prove spectral localization (\autoref{main_result}), it is enough to prove \autoref{main_theorem_reform}. That is, we need to show for almost every $\omega$, for every generalized eigenvalue $E$ of $H_\omega$, there is some $C>0$ such that for all $n$ sufficiently large, $2 n$ and $2 n+ 1$ are $(C, n, E, \omega)$-regular.

The strategy of this proof is as follows:
\begin{enumerate}
	\item The main idea of the proof is to show that, far enough away from a singular point, we have regularity. We will make the argument that, for a generalized eigenfunction $\psi \neq 0$, that we can take 0 to be that singular point and far enough away we will see the desired exponential decay. 
	\item We do this by a contradiction argument--suppose that $2n + 1$ is singular for arbitrarily large $n$. This would correspond to the generalized eigenvalue being in the set of large deviations. However, these sets are little neighborhoods of roots of our characteristic polynomials, hence we get two eigenvalues that are arbitrarily close together.
	\item We use the closeness of these eigenvalues to argue that some entry of the Green's function is bounded below by an exponential term. It is this bound that generates the contradiction that we seek.
	\item However, to get this contradiction, we break into three cases, depending on how close this ``large entry'' is to the edges of the Green's function matrix:
	\begin{enumerate}[leftmargin=45pt]
		\item[\textbf{Case 1:}] We are close to the middle of the matrix and far away from the edges.
		\item[\textbf{Case 2:}] We are close to the edge of the matrix but away from the corner.
		\item[\textbf{Case 3:}] We are close to the corner of the matrix.
	\end{enumerate}
	We only need consider these cases since the Green's function matrix is symmetric. In each case, we will derive a contradiction for $n$ sufficiently large.
	\end{enumerate}
	\begin{figure}[ht]
	\begin{tikzpicture}[scale=1,
            rednode/.style={circle, draw=red!60, fill=red!5, very thick, minimum size=5mm},
            greennode/.style={circle, draw=fgreen!60, fill=fgreen!5, very thick, minimum size=5mm},
            bluenode/.style={circle, draw=blue!60, fill=blue!5, very thick, minimum size=5mm}
            ]
		\filldraw[blue!10] (-2.8, 2.8) -- (1, 2.8) -- (1, 1) -- (-1, 1) -- (-2.8, 2.8);
		\filldraw[blue!10] (2.8, -2.8) -- (2.8, 1) -- (1, 1) -- (1, -1) -- (2.8, -2.8);
		\filldraw[fgreen!10] (1, 1) -- (1, 2.8) -- (2.8, 2.8) -- (2.8, 1) -- (1, 1);
		\filldraw[red!10] (1, 1) -- (1, -1) -- (-1, 1) -- (1, 1);
		\draw[thick] (-2.8, -3) -- (-3, -3) -- (-3, 3) -- (-2.8, 3);
		\draw[thick] (2.8, -3) -- (3, -3) -- (3, 3) -- (2.8, 3);
		\draw[thick, dashed] (-2.8, 2.8) -- (2.8, -2.8) ;
		\node[anchor=south] (c-n) at (-3, 3) {$-n$};
		\node[anchor=south] at (-1.5, 3) {$\cdots$};
		\node[anchor=south] at (0, 3) {0};
		\node[anchor=south] at (1.5, 3) {$\cdots$};
		\node[anchor=south] (cn) at (3, 3) {$n$};
		\node[anchor=east] (r-n) at (-3, 3) {$-n$};
		\node[anchor=east] at (-3, 1.5) {$\vdots$};
		\node[anchor=east] at (-3, 0) {0};
		\node[anchor=east] at (-3, -1.5) {$\vdots$};
		\node[anchor=east] (rn) at (-3, -3) {$n$};
		\draw[thick] (-2.8, 1) -- (2.8, 1);
		\draw[thick] (1, -2.8) -- (1, 2.8);
		\draw [
            thick,
            decoration={
                brace,
                mirror,
                raise=0.2cm
            },
            decorate
        ] (-2.8, -3) -- (0.8, -3) 
        node [pos=0.5,anchor=north,yshift=-0.25cm] {$|n-y_2|\geq \ln(n^p)$}; 
        \draw [
            thick,
            decoration={
                brace,
                mirror,
                raise=0.2cm
            },
            decorate
        ] (1.2, -3) -- (2.8, -3) 
        node [pos=0.5,anchor=north,yshift=-0.25cm] {$|n-y_2|\leq \ln(n^p)$}; 
        \draw [
            thick,
            decoration={
                brace,
                mirror,
                raise=0.2cm
            },
            decorate
        ] (3, -2.8) -- (3, 0.8) 
        node [pos=0.6,anchor=north, xshift=2cm] {$|-n-y_1|\geq \ln(n^p)$}; 
        \draw [
            thick,
            decoration={
                brace,
                mirror,
                raise=0.2cm
            },
            decorate
        ] (3, 1.2) -- (3, 2.8) 
        node [pos=0.7,anchor=north, xshift=2cm] {$|-n-y_1|\leq \ln(n^p)$}; 
        \node[rednode] (1) at (0.35, 0.35) {1};
        \node[greennode] (3) at (2, 2) {3};
        \node[bluenode] (2a) at (2, 0) {2};
        \node[bluenode] (2a) at (0, 2) {2};
	\end{tikzpicture}
	\caption{A skematic of the three cases in the proof of \autoref{main_theorem_reform}.}
	\end{figure}
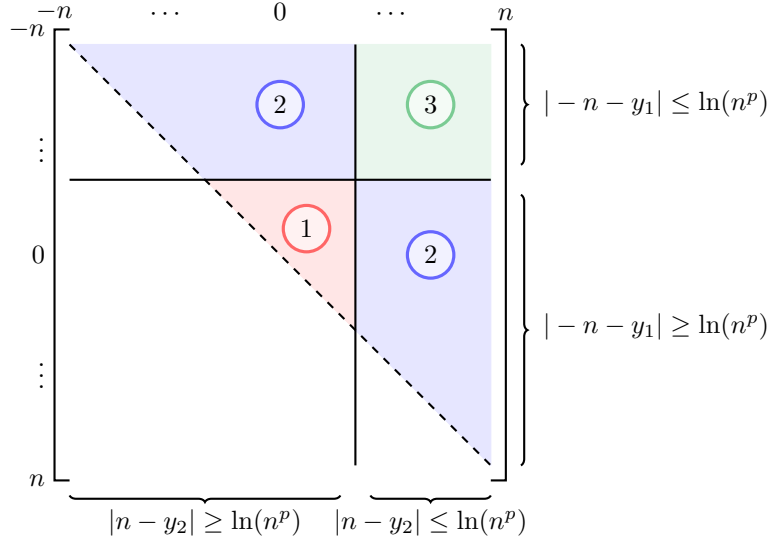

Let us now proceed with the formal proof.

\begin{proof}[Proof of \autoref{main_theorem_reform}]
Fix $\epsilon >0$ and let $\delta >0$ be the corresponding deviation parameter from \autoref{LDT_for_P}. We will show that, for sufficiently large $n$, $2n + 1$ is $(C, n, \tilde{E}, \omega)$-regular with $C = h - 6\epsilon$. The proof for $2n$ is similar. 

For the sake of our technical lemmas, set $r > 1$, $0< \eta < \delta$, $0< \epsilon_0 < \min\{(\delta - \eta)/5, \epsilon\}$, and $p > 6/\delta_0$. Here, $\delta_0$ is the corresponding deviation parameter for $\epsilon_0$, $\eta$ is as in \autoref{tech_lem_2}, and $r$ and $p$ are as in \autoref{tech_lem_3}, \autoref{tech_lem_4}, and \autoref{tech_lem_5}. Obtain a full $\P$-measure set $\Omega_0$ such that every $\omega \in \Omega_0$ the conclusions of \autoref{max_corr}, \autoref{growth_funct_rel_2}, \autoref{tech_lem_1}, \autoref{tech_cor_1}, \autoref{tech_lem_2}, \autoref{tech_lem_3}, \autoref{tech_lem_4}, and \autoref{tech_lem_5} hold. Fix $\omega \in \Omega_0$ and set $N$ sufficiently large as required for these lemmas.

Let $\tilde{E} \in J$ be a generalized eigenvalue for $H_\omega$ with generalized eigenfunction $\psi$. Assume, WLOG, that $\psi(0) \neq 0$. We argue that 0 is $(h - 6 \epsilon, n, \tilde{E}, \omega)$-singular by contradiction. Hence we suppose, for the moment, that 0 is $(h - 6 \epsilon, n, \tilde{E}, \omega)$-regular. Setting $x = 0$, $a = -n$, $b = n$, $E = \tilde{E}$ in \eqref{eq:green-to-ev}, we have
\begin{equation}\label{eq:0_sing}
\psi(0)=-G_{[-n, n], \tilde{E}, \tilde{\omega}}(0, -n) \psi(-n-1)-G_{[-n, n], \tilde{E},\omega}(0, n) \psi(n+1).
\end{equation}
If 0 was $\left(h - 6 \epsilon, n, \tilde{E}, \omega\right)$-regular we would have
\[\abs{G_{[-n,n], \tilde{E}, \tilde{\omega}}(0, -n)} \leq e^{-(h - 6 \epsilon)n} \word{and} \abs{G_{[-n,n], \tilde{E}, \tilde{\omega}}(0, n)} \leq e^{-(h - 6 \epsilon)n}.\]
Note that both are decaying exponentially. Also, we supposed that $\psi$ is a generalized eigenfunction, hence it is polynomially bounded. Combined with \eqref{eq:0_sing}, this says that $0 \neq \psi(0) = O(e^{-cn})$, contradiction. So the supposition that 0 was a regular point must have been false.

Now, seeking a contradiction, suppose for infinitely many $n$, $2n + 1$ is $(h - 6 \epsilon, n, \tilde{E}, \omega)$-singular. By \autoref{tech_cor_1}, $\tilde{E} \in B^{-}_{[n+1, 3n + 1], \epsilon, \omega} \subseteq B^{-}_{[n+1, 3n + 1], \epsilon_0, \omega}$. By \autoref{close_to_eigenvalues} 
\[B^{-}_{[n+1, 3n + 1], \epsilon_0, \omega} = \cparen{E : \abs{P_{[n+1,3n+1], E, \omega}} \leq e^{L_{[n+1, 3n+1], E} - (2n + 1)\epsilon_0}}\]
consists of at most $2n +1$ intervals about the roots of $P_{[n+1,3n+1], E, \omega}$. Therefore $\tilde{E}$ must lie in one of these intervals. Let $E_j$ be the root (i.e., eigenvalue of $H_{[n+1,3n+1],\omega}$) closest to $\tilde{E}$. By \autoref{tech_lem_2}, we know $\abs{E_j - \tilde{E}} < m\paren{B^{-}_{[n+1, 3n + 1], \epsilon_0, \omega}} \leq e^{-(\delta - \eta)(2n + 1)}$.

Applying the above argument to 0 in place of $2n + 1$ yields an eigenvalue $E_i$ of $H_{\omega, [-n, n]}$ where $|\tilde{E} - E_i| \leq e^{-(\delta - \eta)(2n + 1)}$. Thus
\[|E_j - E_i| \leq 2e^{-(\delta - \eta)(2n + 1)}.\]
Recalling a fact of matrix norms
\[\norm{(A - \lambda I)^{-1}} = \sup_{\mu \in \sigma(A)} \frac{1}{\abs{\mu - \lambda}}.\]
By setting $\lambda = E_j$, $\mu = E_i$, we have
\[\norm{G_{[-n,n], E_{j}, \omega}} \geq \frac{e^{(\delta - \eta)(2n + 1)}}{2}.\]
Thus, since $G_{[-n,n], E_{j}, \omega}$ is $(2n + 1) \times (2n + 1)$ dimensional, there exists $y_1, y_2 \in [-n, n]$ such that 
\[\abs{G_{[-n,n], E_j, \omega} (y_1, y_2)} \geq \frac{1}{\sqrt{2n + 1}} \cdot \frac{e^{(\delta - \eta)(2n + 1)}}{2}\]
by using the Frobenius norm. Since $G_{[-n,n], E_{j}, \omega}$ is symmetric, take $y_1 \leq y_2$. Applying \eqref{G_to_P_eq} we get
\begin{equation}\label{eq:main_proof_1}
\frac{\abs{P_{[-n, y_1 - 1], E_j, \omega} P_{[y_2 + 1, n], E_j, \omega}}}{\abs{P_{[-n,n], E_j, \omega}}} \geq \frac{1}{\sqrt{2n + 1}} \cdot \frac{e^{(\delta - \eta)(2n + 1)}}{2}.
\end{equation}
By \autoref{tech_lem_3} (with $y = n$), $E_j \not \in B_{[-n,n], \epsilon_0, \omega}$. Therefore 
\begin{equation}\label{eq:main_proof_2}
\abs{P_{[-n,n], E_j, \omega}} \geq e^{-(2n+1)\epsilon_0 + L_{[-n,n], E_j}}.
\end{equation}
Combining \eqref{eq:main_proof_1} and \eqref{eq:main_proof_2} yields
\begin{equation}\label{main_proof_ineq}
\abs{P_{[-n, y_1 - 1], E_j, \omega}} \cdot \abs{P_{[y_2 + 1, n], E_j, \omega}} \geq \frac{1}{2\sqrt{2n + 1}}e^{(\delta - \eta)(2n + 1) -(2n+1)\epsilon_0 + L_{[-n,n], E_j}}.
\end{equation}
To leverage all our lemmas, we now consider three cases:
\begin{align*}
	& \text{\textbf{Case 1:} $|-n -y_1| \geq \ln(n^p)$ and $|n - y_2| \geq \ln(n^p)$}\\
	& \text{\textbf{Case 2:} $|-n -y_1| \geq \ln(n^p)$ and $|n - y_2| \leq \ln(n^p)$}\\
	& \text{\textbf{Case 3:} $|-n -y_1| \leq \ln(n^p)$ and $|n - y_2| \leq \ln(n^p)$}
\end{align*}
The argument for the case when $|-n -y_1| \leq \ln(n^p)$ and $|n - y_2| \geq \ln(n^p)$ is the same as the argument for Case 2. In each case we derive a contradiction.

\textbf{Case 1:} By applying \autoref{tech_lem_3} to $\abs{P_{[-n, y_1 - 1], E_j, \omega}}$ and $\abs{P_{[y_2 + 1, n], E_j, \omega}}$ we have
\[\abs{P_{[-n,y_1 - 1], E_j, \omega}} \leq e^{(y_1 + n)\epsilon_0 + L_{[-n,y_1 - 1], E_j}}\word{and}\abs{P_{[y_2 + 1, n], E_j, \omega}} \leq e^{(n - y_2)\epsilon_0 + L_{[y_2 + 1, n], E_j}}\]
hence, using \eqref{main_proof_ineq}, 
\begin{align*}
	e^{(y_1 + n)\epsilon_0 + L_{[-n,y_1 - 1], E_j}}  e^{(n - y_2)\epsilon_0 + L_{[y_2 + 1, n], E_j}} & \geq \frac{1}{2\sqrt{2n + 1}}e^{(\delta - \eta)(2n + 1) -(2n+1)\epsilon_0 + L_{[-n,n], E_j}}\\
	e^{(2n + y_1 - y_2)\epsilon_0 + L_{[-n,y_1 - 1], E_j}+ L_{[y_2 + 1, n], E_j}} & \geq \frac{1}{2\sqrt{2n + 1}}e^{(\delta - \eta)(2n + 1) -(2n+1)\epsilon_0 + L_{[-n,n], E_j}}
\end{align*}
and so
\begin{align*}
	\exp\bracket{(4n + y_1 - y_2 + 1)\epsilon_0 - (\delta - \eta)(2n + 1) + L_{[-n,y_1 - 1], E_j}+ L_{[y_2 + 1, n], E_j} - L_{[-n,n], E_j}}\\*
	 \qquad  \geq \frac{1}{2\sqrt{2n + 1}}. \numberthis \label{eq:case1}
\end{align*}
Now, from \autoref{growth_funct_rel_2}, we know that 
\[0 \leq L_{[-n, y_1 - 1], E_j} + L_{[y_1 - 1, n], E_j} - L_{[-n, n], E_j} \leq 2n\epsilon_0\]
and
\[0 \leq L_{[y_1 - 1, y_2 + 1], E_j} + L_{[y_2 + 1, n], E_j} - L_{[y_1 - 1, n], E_j} \leq 2n\epsilon_0\]
adding these together yields
\[0 \leq L_{[-n, y_1 - 1], E_j} + L_{[y_1 - 1, y_2 + 1], E_j} + L_{[y_2 + 1, n], E_j} - L_{[-n, n], E_j} \leq 4n\epsilon_0. \]
Invoking the fact that $L_{[y_1 - 1, y_2 + 1], E_j}$ is positive, we get 
\[L_{[-n, y_1 - 1], E_j} + L_{[y_2 + 1, n], E_j} - L_{[-n, n], E_j} \leq 4n\epsilon_0. \]
Therefore \eqref{eq:case1} becomes
\begin{align*}
	\exp\bracket{(8n + y_1 - y_2 + 1)\epsilon_0 - (\delta - \eta)(2n + 1)} & \geq \frac{1}{2\sqrt{2n + 1}}\\
	\exp\bracket{(8n + 2)\epsilon_0 - (\delta - \eta)(2n + 1)} & \geq \frac{1}{2\sqrt{2n + 1}}\\
\end{align*}
which yields a contradiction since the LHS decays to 0 as $n \to \infty$ faster than the RHS since $\epsilon_0 \leq (\delta - \eta)/5$.

\textbf{Case 2:} By applying \autoref{tech_lem_3} to $\abs{P_{[-n, y_1 - 1], E_j, \omega}}$ and \autoref{tech_lem_5} to $\abs{P_{[y_2 + 1, n], E_j, \omega}}$ we have
\[\abs{P_{[-n,y_1 - 1], E_j, \omega}} \leq e^{(y_1 + n)\epsilon_0 + L_{[-n,y_1 - 1], E_j}}\word{and}\abs{P_{[y_2 + 1, n], E_j, \omega}} \leq n^{p(\ln(n^p) + 1)+ r/\gamma}\]
hence, using \eqref{main_proof_ineq},
\begin{align*}
	e^{(y_1 + n)\epsilon_0 + L_{[-n,y_1 - 1], E_j}} \cdot n^{p(\ln(n^p) + 1)+ r/\gamma} & \geq \frac{1}{2\sqrt{2n + 1}}e^{(\delta - \eta)(2n + 1) -(2n+1)\epsilon_0 + L_{[-n,n], E_j}}
\end{align*}
and so
\begin{align*}
	 \exp\bracket{(\eta - \delta)(2n+1) + (3n + y_1 + 1)\epsilon_0 - L_{[-n, n], E_j} + L_{[-n, y_1  - 1], E_j}}\cdot n^{p(\ln(n^p) + 1)+ r/\gamma}\\
	  \qquad  \geq \frac{1}{2\sqrt{2n + 1}}. \numberthis \label{eq:case2}
\end{align*}
Invoking the work above, we have $L_{[-n, y_1 -1], E_j} - L_{[-n,n], E_j} \leq 4 n \epsilon_0$ and so \eqref{eq:case2} becomes 
\[\exp\bracket{4(2n + 1) \epsilon_0 - (\delta - \eta)(2n + 1)}\cdot  n^{p(\ln(n^p) + 1)+ r/\gamma} \geq \frac{1}{2\sqrt{2n + 1}}.\]
Since $4(2n + 1) \epsilon_0 - (\delta - \eta)(2n + 1) < 0$ by assumption, the LHS decays to 0 as $n \to \infty$ faster than the RHS, which again yields a contradiction.

\textbf{Case 3:} Invoking \autoref{tech_lem_5} twice gives us
\[\abs{P_{[-n,y_1 - 1], E_j, \omega}}\cdot \abs{P_{[y_2 + 1, n], E_j, \omega}} \leq n^{2p(\ln(n^p) + 1)+ 2r/\gamma}.\]
Applying this to \eqref{main_proof_ineq},
\begin{align*}
	\frac{1}{2\sqrt{2n + 1}} & \leq  n^{2p(\ln(n^p) + 1)+ 2r/\gamma} \cdot \exp\bracket{(2n + 1)\epsilon_0 - (2n + 1)(\delta - \eta) - L_{[-n,n], E_j}}\\
	& \leq  n^{2p(\ln(n^p) + 1)+ 2r/\gamma} \cdot \exp\bracket{(2n + 1)\epsilon_0 - (2n + 1)(\delta - \eta) - n\cdot h}
\end{align*}
where $h$ comes from \autoref{noniidFurst}. Again, our assumptions imply $(2n + 1)\epsilon_0 - (2n + 1)(\delta - \eta) - n \cdot h < 0$ and so we once again have a contradiction as the RHS decays to 0 as $n \to \infty$ faster than the LHS. 

In all cases, we have a contradiction and so $2n + 1$ is eventually $(h - 6\epsilon, n, \tilde{E}, \omega)$-regular and the proof is complete.
\end{proof}


\section{Proof of Dynamical Localization} \label{Dynamical Localization}

Now that we have established spectral localization, we can also establish dynamical localization with some additional work by adapting the ideas above. We will actually establish the stronger property, semi-uniformly localized eigenfunctions:
\begin{definition}
A self-adjoint operator $H: \ell^2(\mathbb{Z}) \to \ell^2(\mathbb{Z})$ has semi-uniformly localized eigenfunctions (SULE) if $H$ has a complete set $\left\{\psi_E\right\}$ of orthonormal eigenfunctions (where $\psi_E$ denotes the eigenfunction with eigenvalue $E$), and there is $\alpha>0$ such that for each $\xi>0$ there exists a constant $C_{\xi}$ so that for any eigenvalue $E$ there exists $l_E \in \Z$ such that
\[
\left|\psi_E(m)\right| \leq C_{\xi} e^{\xi\left|l_E\right|-\alpha\left|m-l_E\right|}
\]
for all $m \in \mathbb{Z}$.
\end{definition}

By \cite{Rio:1996aa}, establishing SULE will provide dynamical localization. In fact, we will prove that 
\begin{equation}\label{dynam_loc_goal}
\left|\psi_E(x)\right| \leqslant \tilde{C} e^{C \ln ^2\left(1+\left|l_E\right|\right)} e^{-\alpha\left|x-l_E\right|}.
\end{equation}
This will be accomplished by utilizing the work done above; we will leverage the proof in Section \ref{Proof of Main Theorem} to achieve a more general statement that allows for viewing other windows besides the one centered at 0. We also want to make sure that the large $N$, for which our technical lemmas in Sections \ref{Properties} and \ref{Technical Lemmas} hold, is uniform in $E \in J$. All except \autoref{max_corr} already satisfy this. To get a statement like \autoref{max_corr} uniform in $E\in J$, we require a one-sided version of the LDE which is uniform in $E \in J$. This is achieved through the following consequence of equicontinuity of $\cparen{\frac{1}{n}L_{n, E}}$:

\begin{lemma}\label{unif_LDE}
For any $\bar{\epsilon}>0$ there exists $c_2, C_2>0$ and $N' \in \mathbb{N}$ such that for any $n>N'$ with the probability at least $1-C_2\exp \left(-c_2 n\right)$ the following statement holds: for any $E \in J$ one has
\[
\log \left\|T_{n, E, \omega}\right\|-L_{n,E} \leq n \bar{\epsilon} .
\]
\end{lemma}

\begin{proof}
Fix $\epsilon >0$ and take $\epsilon^*$, $\delta^*$, $k$, and $\delta$ be as in the proof of \autoref{prelim_equicont}. Take points $\{E_1,...,E_N\} \subseteq J$ such that $J$ is divided into intervals of length less than $\delta$. Notice that $N$ does not depend on $m = \frac{n}{k}$ since $\delta$ does not depend on $m$. Under the same argument as for \eqref{eq:equicont4} in \autoref{prelim_equicont}, for all $1 \leq i \leq N$ we have that all the inequalities 
\begin{equation}\label{unif_LDE_eq1}
\left|\log \left\|T_{n, E_i, \omega}\right\|-\left(\xi_{1, E_i}+\cdots+\xi_{m, E_i}\right)\right|<\frac{\varepsilon}{5} n
\end{equation}
hold with probability at least $1 - Ne^{-\delta^*m}$. Thus we have that
\begin{align*}
	& \abs{L_{n, E_i} - \paren{\xi_{1, E_i}+\cdots+\xi_{m, E_i}}}\\ 
	& \leq \abs{L_{n, E_i} - \log\norm{T_{n, E_i, \omega}}} + \abs{\log\norm{T_{n, E_i, \omega}} - \paren{\xi_{1, E_i}+\cdots+\xi_{m, E_i}}}\\
	& < \frac{\epsilon}{5}n + \frac{\epsilon}{5}n \numberthis \label{unif_LDE_eq2}
\end{align*}
holds simultaneously for all $1 \leq i \leq N$ with probability at least $1 - 2Ne^{-\delta^*m}$ by \autoref{paraFurst} and \eqref{unif_LDE_eq1}.

Now take $E \in J$ to be arbitrary and let $E_i \in \{E_1,...,E_N\}$ be such that $|E - E_i| < \delta$. By the equicontinuity of $\xi_{j, E}$ as proved in \autoref{prelim_equicont}, we have
\begin{align*}
	\log\norm{T_{n, E, \omega}} & \leq \paren{\xi_{1, E}+\cdots+\xi_{m, E}}\\
	& \leq \paren{\xi_{1, E_i}+\cdots+\xi_{m, E_i}} + \frac{\epsilon}{5}n.
\end{align*}
Using this in conjunction with \eqref{unif_LDE_eq2} we have
\[\log\norm{T_{n, E, \omega}} \leq L_{n, E_i} + \frac{3}{5}\cdot \epsilon\cdot n.\]
Invoking the result of \autoref{prelim_equicont} yields
\[\log\norm{T_{n, E, \omega}} \leq L_{n, E} + \frac{8}{5}\cdot \epsilon\cdot n\]
with probability $1 - 2Ne^{-\delta^*m}$. Setting $\bar{\epsilon} = \frac{8}{5} \epsilon$, $C_2 = 2N$, $c_2 = \delta^*$ finishes the proof.
\end{proof}

This result extends to characteristic polynomials as well, since
\begin{align*}
	\log\abs{P_{[1,n], E, \omega}} - L_{n, E} & = \log\abs{\langle T_{n, E, \omega} e_1, e_1 \rangle} - L_{n, E}\\
	& \leq \log\norm{T_{n, E, \omega} e_1}^{1/2} - L_{n, E}\\
	& \leq \log\norm{T_{n, E, \omega}} - L_{n, E}\\
	& < \bar{\epsilon}n
\end{align*}
Using this in place of \autoref{LDT_for_P}, the proof of the following corollary runs the same as the proof of \autoref{max_corr}.

\begin{corollary}\label{unif_max_corr}
For almost every $\omega$, for every $\epsilon >0$ there is an $N(\omega, \epsilon) = N \in \N$ sufficiently large such for all $n \geq N$ and all $E \in J$ we have
\[\max\cparen{\log\abs{P_{[n+1,2n], E, \omega}} - L_{[n+1,2n], E, \omega}, \log\abs{P_{[2n+2,3n+1], E, \omega}} - L_{[2n+2,3n+1], E, \omega}} < \epsilon n.\]
\end{corollary}

Using this, we are able to extend the proof of spectral localization in Section \ref{Proof of Main Theorem} to more general windows and $N$ uniform in $E \in J$:

\begin{lemma} \label{sing_implies_far_reg}
There exists a full-measure set $\Omega_1$ such that for any $l$ and $\omega \in \Omega_1$, there exists $N(l, \omega)$, such that for any $n>N(l, \omega)$ and for all $E \in \sigma(H_\omega)$ either $l$ or $l\pm(2 n+1)$ is $(h - 6 \epsilon, n, E, \omega)$-regular.
\end{lemma}

\begin{proof}
In the proof of \autoref{main_theorem_reform} in Section \ref{Proof of Main Theorem} we showed that for every generalized eigenvalue $E$ of $H_\omega$, in the case when 0 is $(h - 6 \epsilon, n, E, \omega)$-singular we have that $2n+1$ must be $(h - 6 \epsilon, n, E, \omega)$-regular for all $n>N(\omega)$, with $N(\omega)$ sufficiently large. Let $T$ be the shift operator on the sequence space $\Omega$ and set $N(l, \omega)=\max \left\{N(T^l(\omega)), N(T^{-l}(\omega))\right\}$. Repeat the proof in Section \ref{Proof of Main Theorem} with $l$ in place of 0. Then we may use $\Omega_1 = \bigcap_{l \in \mathbb{Z}} T^l (\Omega_0)$, where $\Omega_0$ is the full-measure set as in the proof of \autoref{main_theorem_reform} in Section \ref{Proof of Main Theorem}.
\end{proof}

The next lemma will allow us to get the $\ln ^2\left(1+\left|l_E\right|\right)$ term in the exponent in \eqref{dynam_loc_goal}. First, some notation: let $N_1(\omega)$ be as from \autoref{tech_lem_2} and, similarly, set $N_2(\omega)$ to the be the smallest index such that $n > N_2(\omega)$ implies $\omega \not \in C_n \cup D_n$ as in \autoref{tech_lem_3}. 

\begin{lemma} \label{dynam_loc_tech_lem_1} 
There exists a full-measure set $\Omega_2$, such that  for any $\omega \in \Omega_2$ there is $L(\omega)$ such that for any $|l| > L(\omega)$ we have
\[ \max \left\{N_1(l, \omega), \ N_2(l, \omega)\right\} \leqslant \ln ^2|l|\]
where $N_i(l, \omega) = N_i(T^l(\omega))$.
\end{lemma}

\begin{proof}
We show this for $N_1(l,\omega)$ first; the proof for $N_2(l,\omega)$ is similar. Let $\omega \in \Omega_1$, where $\Omega_1$ is as in \autoref{sing_implies_far_reg}. We have
\begin{align*}
	& \P\cparen{N_1(l,\omega) > k}\\ 
	& \leq \sum_{n=k+1}^\infty \P\cparen{N_1(l,\omega) = n}\\
	& \leq \sum_{n=k+1}^\infty \P\cparen{\max \cparen{m\paren{B^-_{[n+1, 3n+1], \epsilon, \omega}}, \ m\paren{B^-_{[-n, n], \epsilon, \omega}}} > e^{-(\delta - \eta)(2n+1)}} \\
	& \leq \sum_{n=k+1}^\infty 2m(J)e^{-\eta(2n+1)}\numberthis\label{eq:dynam_loc_lem1}\\
	& \leq C e^{-\eta(2k+1)}
\end{align*}
where line \eqref{eq:dynam_loc_lem1} comes from the proof of \autoref{tech_lem_2}. Therefore
\[\P\cparen{N_1(l,\omega) > \ln^2|l|} \leq C e^{-\eta(2\ln^2|l|+1)}\]
The result now follows from an application of Borel-Cantelli.
\end{proof}

Now we may prove dynamical localization by demonstrating \eqref{dynam_loc_goal}. The strategy is as follows:
\begin{enumerate}
	\item We first isolate where a maximum occurs in the eigenfunction $\psi_E$ and denote this index as $l_E$. This index plays the role of 0 in the proof of spectral localization; that is, it will be the singular point which will be the center of our windows.
	\item We consider entries $x$ of $\psi_E$ defined by $x = l_E \pm (2n + 1)$, with the treatment of entries of the form $l_E \pm 2n$ being similar. We wish to show that all entries $x$ obey \eqref{dynam_loc_goal}, so we split our analysis into two cases: when $n$ is sufficiently large so that $x$ is a regular point and when it is not. 
	\begin{enumerate}
		\item The ``$x$ is regular'' case can be shown to obey \eqref{dynam_loc_goal} with minimal manipulation.
		\item The other case requires more massaging. We consider two scenarios; when the maximum point $l_E$ is far away from 0 and when it is close to 0 ($|l_E| \geq L$ and $|l_E| < L$ in the proof, respectively). When $l_E$ is ``far away,'' we use \autoref{dynam_loc_tech_lem_1} to get the desired \eqref{dynam_loc_goal}. When $l_E$ is ``close,'' we simply deal with the finite number of entries that this case covers by adjusting the constant $C_\epsilon$.
	\end{enumerate} 
\end{enumerate}

\begin{proof}[Proof of \autoref{dynam_loc}] 
Take $\tilde{\Omega} = \Omega_1 \cap \Omega_2$, which is full-measure, and fix $\omega \in \tilde{\Omega}$. We now omit $\omega$ from the notation. Set $L$ as in \autoref{dynam_loc_tech_lem_1} and $N(l) = \max\{N_1(l), N_2(l)\}$.

Let $l_E$ be position of the maximum point of $\abs{\psi_E}$. Without loss of generality, normalize $\psi_E$ so that $\norm{\psi_E} = 1$. Note that for $n > \frac{ln(2)}{h - 6 \epsilon}$, $l_E$ is $(h - 6 \epsilon, n, E, \omega)$-singular. To see this, suppose $l_E$ was $(h - 6 \epsilon, n, E, \omega)$-regular. Then
\[\abs{G_{[l_E - n, l_E + n], E, \omega} (l_E, l_E \pm n)} < \exp\bracket{-(h - 6 \epsilon)\paren{\frac{\ln(2)}{h - 6 \epsilon}}} = \frac{1}{2}\]
Therefore, by \eqref{eq:green-to-ev}, we have
\begin{align*}
	\abs{\psi_E(l_E)} & = |G_{[l_E - n, l_E + n], E, \omega} (l_E, l_E - n)\cdot \psi_E(l_E - n - 1)\\*
	& \quad + G_{[l_E - n, l_E + n], E, \omega} (l_E, l_E + n) \cdot \psi_E(l_E + n + 1)|\\
	& < \frac{1}{2}\paren{\abs{\psi_E(l_E - n - 1) } + \abs{\psi_E(l_E + n + 1) }}
\end{align*}
impossible since $l_E$ is the maximum point, hence $l_E$ is $(h - 6 \epsilon, n, E, \omega)$-singular.

Now, consider the case when $\abs{x - l_E} \geq N(l_E)$. By the above and \autoref{sing_implies_far_reg}, $x = l_E \pm (2n + 1)$ is $(h - 6 \epsilon, n, E, \omega)$-regular. Again by \eqref{eq:green-to-ev}, we have 
\begin{align*}
	\abs{\psi_E(x)} & = |G_{[x - n, x + n], E, \omega} (x, x - n)\cdot \psi_E(x - n - 1) \\
	& \qquad + G_{[x - n, x + n], E, \omega} (x, x + n) \cdot \psi_E(x + n + 1)|\\
	& \leq \abs{G_{[x - n, x + n], E, \omega} (x, x - n)} + \abs{G_{[x - n, x + n], E, \omega} (x, x + n)}\\
	& \leq 2 e^{-(h-6\epsilon)n}\\
	& = 2e^{-(h-6\epsilon)\bracket{\frac{1}{2}|x - l_E| - \frac{1}{2}}}\\
	& = (2 e^{\frac{1}{2}(h-6\epsilon)})e^{-\frac{1}{2}(h-6\epsilon)|x - l_E|}\\
	& \leq (2 e^{\frac{1}{2}(h-6\epsilon)})e^{-\frac{1}{2}(h-6\epsilon)|x - l_E|}\cdot e^{\frac{1}{2}(h-6\epsilon)\ln^2(|l_E| + 1)} \numberthis \label{eq:dynam_loc_1}
\end{align*}
where the last line follows since $(h-6\epsilon)\ln^2(|l_E| + 1) \geq 0$. 

Now consider the case when $\abs{x - l_E} < N(l_E)$ and $\abs{l_E} \geq L$. Recall $\psi_E$ is normalized so $|\psi_E(x)| \leq 1$ for all $x$. Therefore
\begin{equation*}
\abs{\psi_E(x)} \leq 1 \leq 2e^{-\frac{1}{2}(h-6\epsilon)|x - l_E|} \cdot e^{\frac{1}{2}(h-6\epsilon)N(l_E)}.
\end{equation*}
Invoking \autoref{dynam_loc_tech_lem_1}, we have
\begin{equation}\label{eq:dynam_loc_2}
\abs{\psi_E(x)} \leq 2e^{-\frac{1}{2}(h-6\epsilon)|x - l_E|} \cdot e^{\frac{1}{2}(h-6\epsilon)\ln^2(|l_E| + 1)}.
\end{equation}

Lastly, consider the case when $\abs{x - l_E} < N(l_E)$ and $\abs{l_E} < L$. Set 
\[M_{\epsilon} = \min_{k \in [-L,L], \ |x-k|<N(k)}\cparen{e^{\frac{1}{2}(h-6\epsilon) \ln^2(k+1)}e^{-\frac{1}{2}(h-6\epsilon)|x-k|}}.\]
With $C_\epsilon = M_\epsilon^{-1}$, we have for all $|x - l_E| < N(l_E)$
\begin{equation}\label{eq:dynam_loc_3}
|\psi_E(x)| \leq 1 \leq C_\epsilon \cdot e^{-\frac{1}{2}(h-6\epsilon)|x - l_E|} \cdot e^{\frac{1}{2}(h-6\epsilon)\ln^2(|l_E| + 1)}.
\end{equation}
Combining estimates \eqref{eq:dynam_loc_1}, \eqref{eq:dynam_loc_2}, and \eqref{eq:dynam_loc_3} and setting $\tilde{C} = \max\{C_\epsilon, 2 e^{\frac{1}{2}(h-6\epsilon)},2\}$, $\alpha = C = \frac{1}{2}(h-6\epsilon)$ yields SULE and hence dynamical localization.
\end{proof}


\appendix

\section{Necessity of the Main Assumptions}\label{assumptions}

The second assumption in \autoref{main_result} is there to ensure that, in the limit, no distributions are deterministic. If there was a subsequence of potentials that converged to a deterministic distribution, then for our compact set of probability measures $\mathcal{K}$ the ``measures condition'' of \autoref{noniidFurst} fails and we cannot apply the Non-Stationary Furstenberg Theorem.

However, there is a difference in the $0< \gamma \leq 2$ and the $\gamma > 2$ regimes in terms of what we require of the variances of our potentials. When $\gamma > 2$, we may take the naive hypothesis that, uniformly, each potential has variation bounded away from 0. However, when $0< \gamma \leq 2$ this is not sufficient. In this section, we will discuss the precise differences in these $\gamma$-regimes.

\textbf{When }$\mathbf{0< \gamma \leq 2}$\textbf{:} In the case where $0< \gamma \leq 2$, our ``no deterministic distributions'' condition aims to encode the requirement that a non-trivial amount of variance of every potential must live in some common compact interval. We do this by considering $\mathrm{Var}(\truncated{V_\omega(n)})$, which is the potential $V_\omega(n)$ truncated to $[-k,k]$ with the mass outside of this interval sent to $-k$ and $k$, whichever is closer. As the next lemma shows, requiring that there exists some $\epsilon> 0$ such that 
\[\mathrm{Var}(\truncated{V_\omega(n)}) > \epsilon\]
for all $n$ yields non-determinism of each $V_\omega(n)$.

\begin{lemma}\label{non-deterministic_lemma}
Let $k > 0$. For a random variable $X$, if $\truncated{X}$ is not a.s. constant, then neither is $X$.
\end{lemma}

\begin{proof}
We proceed by contrapositive. Suppose $X$ was a.s. constant, i.e., $X = c$ with probability 1. Then
\[\truncated{X} = \begin{cases}
-k & \text{if }c < -k \\
c & \text{if }-k \leq c \leq k\\
k & \text{if }k < c\\
\end{cases}\]
Since $c$ is a constant, $\truncated{X}$ falls into one of these possibilities with probability 1, and hence is a.s. constant.
\end{proof}

The reader may wonder why it is necessary to consider the more complicated requirement that $\mathrm{Var}(\truncated{V_\omega(n)}) > \epsilon$. The problem is that, under the weaker but simpler requirement that $\mathrm{Var}(V_\omega(n)) > \epsilon$, the sequence of distributions may still converge to a deterministic measure in the weak limit (see Example \ref{deterministic_limit_ex} for an explicit example). However, if we impose the assumption that $\mathrm{Var}(\truncated{V_\omega(n)}) > \epsilon$ then nontrivial variance is preserved under the weak limit. Formally:

\begin{lemma}\label{limit_of_variances_lemma}
Let $(V_\omega(n))_n$ be a sequence of independent random variables. Assume that there exists $\epsilon, k > 0$ such that $\mathrm{Var}(\truncated{V_\omega(n)}) > \epsilon$ for all $n$. If the distribution of $\truncated{V_\omega(n)}$ converges weakly, the limiting distribution also has variance bounded from below by $\epsilon$.
\end{lemma}

\begin{proof} Set $W(n,k) := \truncated{V_\omega(n)}$ to make the notation simpler. Assume $\mathrm{Var}(W(n,k)) > \epsilon$ for all $n$. Also assume that $\{V_\omega(n)\}_{n}$, and hence $\{W(n,k)\}_{n}$, is weakly converging as $n \to \infty$. We would like to demonstrate that the limit of variances $\mathrm{Var}(W(n,k))$ is the variance of the limiting distribution. This will allow us to conclude that the limiting distribution has variance also bounded below by $\epsilon$. 

Since $\mathrm{Var}(X) = \E[X^2] - \E[X]^2$, it is enough to show that $W(n,k)$ converges in $\mathcal{L}^1$ and $\mathcal{L}^2$. To that end, we will show $W(n,k)$ and $W(n,k)^2$ both converge in $\mathcal{L}^1$. Note $W(n,k) \in [-k,k]$ for all $n$ and thus $|W(n,k)| \leq k$, $W(n,k)^2 \leq k^2$. That is, $\{|W(n,k)|\}_{n}$ and $\{W(n,k)^2\}_{n}$ are both uniformly bounded families of random variables. Uniformly bounded families of random variables are also uniformly integrable\footnote{We say a class $\mathcal{C}$ of random variables is called \textit{uniformly integrable} if given $\varepsilon>0$ there exists $K$ in $[0, \infty)$ such that $\mathrm{E}(|X| ;|X|>K)<\varepsilon$ for all $X \in \mathcal{C}$}. If a family of random variables is uniformly integrable and converges in distribution (i.e., weakly), they also converge in $\mathcal{L}^1$ (see \cite{Billingsley:1999aa}, Theorem 3.5). Hence $\{|W(n,k)|\}_{n}$ and $\{W(n,k)^2\}_{n}$ both converge in $\mathcal{L}^1$ and the variance of the limiting distribution is indeed $\lim_{n \to \infty} \mathrm{Var}(W(n,k)) \geq \epsilon > 0$. \end{proof}

The last thing to check is that if our original sequence $V_\omega(n)$ converges in distribution (i.e., weakly) to some $V(\infty)$ then $V(\infty)$ is not a.s. constant. Indeed, notice that $g(x) = \truncated{x}$ is continuous. The Continuous Mapping Theorem (see Theorem 3.2.10 in \cite{Durrett:2019aa}) says that convergence in distribution is preserved by continuous maps, hence $g(V_\omega(n)) \to g(V(\infty))$ in distribution. That is, $V_\omega(n) \to V(\infty)$ implies $\truncated{V_\omega(n)} \to \truncated{V(\infty)}$. \autoref{limit_of_variances_lemma} implies $\truncated{V(\infty)}$ has nonzero variance and \autoref{non-deterministic_lemma} lets us conclude that $V(\infty)$ is not deterministic. Thus we have shown the following:

\begin{lemma}
Let $V_\omega(n)$ be a sequence of random potentials with distributions $\mu_n$ such that there exists $\epsilon >0$ for which $\var\paren{\truncated{V_\infty(n)}} > \epsilon$ for all $n$. Then each $V_\omega(n)$ is not deterministic and, furthermore, any convergent sub-sequential limit (in distribution) is not deterministic as well.
\end{lemma}

\textbf{When }$\mathbf{\gamma > 2}$\textbf{:} In contrast, the requirement that the variance be nonzero on a compact set is not necessary when $\gamma > 2$. Simply assuming there exists an $\epsilon >0$ such that $\mathrm{Var}(V_\omega(n)) > \epsilon$ is enough. The reason for this is because we can use the finite-gamma moment condition to directly invoke Theorem (a) in section 13.3 of \cite{Williams:2008aa} which says that sequences of random variables which have a uniformly bounded $p$-norm, $p>1$, are uniformly integrable. With $\gamma > 2$, this implies both $\{|V_\omega(n)|\}_n$ and $\{V_\omega(n)^2\}_n$ are uniformly integrable and hence converge in $\mathcal{L}^1$. Without the need to artificially bound our random variables, we get that the limit of the variances is the variance of the limit. Thus the limiting distribution also has nonzero variance. For further reading on uniform integrability and weak convergence, see \cite{Williams:2008aa} Section 13.3 and \cite{Billingsley:1999aa} Section 3.

However, the situation is not as dramatic as it first appears. In fact, the requirement that $\mathrm{Var}(V_\omega(n)) > \epsilon$ for all $n$ in the $\gamma > 2$ regime implies that there is variance that persists in a common compact interval:

\begin{lemma}
Let $V_\omega(n)$ be a sequence of random potentials, each distributed according to $\mu_n$, such that
\begin{enumerate}
	\item there exists $\gamma >2$ and $C_0$ such that for any $n$ we have $\int|x|^\gamma d\mu_n(x) < C_0$, and 
	\item there exists $\epsilon > 0$ such that $\mathrm{Var}(V_\omega(n)) > \epsilon$ for all $n$.
\end{enumerate}
Then there exists $k > 0$ such that $\var(\truncated{V_\omega(n)}) \geq \frac{3\epsilon}{4}$ for all $n$.
\end{lemma}

\begin{proof}
Using the finite $\gamma$-moment condition and Markov's inequality, we know $\P\cparen{|V_\omega(n)|>k} < \frac{C_0}{k^\gamma}$ for all $n$. Setting $W(n,k) = \truncated{V_\omega(n)}$ as above, note 
\begin{align*}
	\E\bracket{V_\omega(n)^2 - W(n,k)^2} & = \E\bracket{\mathbb{1}_{|V_\omega(n)|> k} \cdot (V_\omega(n)^2 - k^2)}\\
	& \leq \E\bracket{\mathbb{1}_{|V_\omega(n)|> k} \cdot V_\omega(n)^2}\\
	& \leq \E\bracket{\abs{V_\omega(n)}^\gamma}^{\frac{2}{\gamma}} \cdot \P\cparen{|V_\omega(n)|> k}^{\frac{\gamma - 2}{\gamma}} & \text{by H\"older's inequality}\\
	& < C_0^{\frac{2}{\gamma}} \cdot \paren{\frac{C_0}{k^\gamma}}^{\frac{\gamma - 2}{\gamma}}\\
	& = \frac{C_0}{k^{\gamma - 2}}
\end{align*}
and
\begin{align*}
	& \abs{\E[V_\omega(n)]^2 - \E[W(n,k)]^2}\\ 
	& = \abs{\E[V_\omega(n) - W(n,k)]\cdot \E[V_\omega(n) + W(n,k)]}\\
	& \leq \abs{\E[V_\omega(n) - W(n,k)]}\paren{\E[|V_\omega(n)|] + \E[|W(n,k)|]}\\
	& \leq \abs{\E[V_\omega(n) - W(n,k)]}\cdot 2C_0^{\frac{1}{\gamma}} \numberthis\label{using_gamma_geq_2}\\
	& = \abs{\E\bracket{\mathbb{1}_{V_\omega(n)> k} \cdot (V_\omega(n) - k) + \mathbb{1}_{V_\omega(n)<- k} \cdot (V_\omega(n) + k)}}\cdot 2C_0^{\frac{1}{\gamma}}\\
	& \leq \paren{\E\bracket{\mathbb{1}_{|V_\omega(n)|> k} \cdot |V_\omega(n)|} + \E\bracket{\mathbb{1}_{|V_\omega(n)|> k} \cdot k}}\cdot 2C_0^{\frac{1}{\gamma}}\\
	& \leq \paren{\E\bracket{\abs{V_\omega(n)}^\gamma}^{\frac{1}{\gamma}} \cdot \P\cparen{|V_\omega(n)|> k}^{\frac{\gamma - 1}{\gamma}} + k \cdot \P\cparen{|V_\omega(n)|> k}^{\frac{\gamma - 1}{\gamma}}}\cdot 2C_0^{\frac{1}{\gamma}}\\
	& < \paren{C_0 \cdot \paren{\frac{C_0}{k^\gamma}}^{\frac{\gamma - 1}{\gamma}} + k \cdot \paren{\frac{C_0}{k^\gamma}}^{\frac{\gamma - 1}{\gamma}}}\cdot 2C_0^{\frac{1}{\gamma}}\\
	& \leq \frac{2C_0^{2}}{k^{\gamma-1}}
\end{align*}
where we use $\gamma >2$ to get $\E[|W(n,k)|] \leq \E[|V_\omega(n)|] \leq \E[|V_\omega(n)|^\gamma]^{1/\gamma} < C_0^{\frac{1}{\gamma}}$ in line \eqref{using_gamma_geq_2}. Putting these together, we may take $k \gg 1$ so that 
\[\abs{\mathrm{Var}(V_\omega(n)) - \mathrm{Var}(W(n,k))} \leq \abs{\E[V_\omega(n)^2] - \E[W(n,k)^2]} + \abs{\E[V_\omega(n)]^2 - \E[W(n,k)]^2} < \frac{\epsilon}{4}\]
whence $\mathrm{Var}(\mathbb{1}_{[-k,k]}(V_\omega(n)) \cdot V_\omega(n)) \geq \mathrm{Var}(V_\omega(n)) - \frac{\epsilon}{4} > \frac{3\epsilon}{4}$ for all $n$.
\end{proof}

\begin{example}\label{deterministic_limit_ex}
Now, as promised, we wish to present an example of why it is not enough to simply assume $\mathrm{Var}(V_\omega(n)) > \epsilon$ in the $0< \gamma \leq 2$ regime. Indeed, consider the example sequence of random variables defined by
\[V_\omega(n) = \begin{cases}
0 & \text{with probability } 1 - \frac{1}{n^2}\\
n & \text{with probability } \frac{1}{n^2}\\
\end{cases}
\]
and denote the distribution as $\mu_n$. Notice that this sequence of random variables obeys the finite $\gamma$-moment condition for $0< \gamma \leq 2$:
\[\E[|V_\omega(n)|^\gamma] = n^\gamma \cdot \frac{1}{n^2} = n^{\gamma - 2} \leq 1\]
for all $n$. Furthermore, $\mathrm{Var}(V_\omega(n)) = 1- \frac{1}{n^2} \geq \frac{3}{4}$ for $n \geq 2$. However, for any $f \in CB(\R)$ (continuous and bounded on $\R$):
\begin{align*}
	\lim_{n \to \infty} \int f(x) d\mu_n(x) & = \lim_{n \to \infty} f(0) \cdot \paren{ 1 - \frac{1}{n^2}} + f(n) \cdot \frac{1}{n^2}\\
	& = f(0)\\
	& = \int f(x) d\delta_0(x)
\end{align*}
where $\delta_0$ is the distribution with a single point mass at 0.

The reader may notice that the variance of $V_\omega(n)$ does converge to 1 as $n \to \infty$ in this example. However, this does not pose a problem for our analysis earlier in this appendix because, even though $\mathrm{Var}(V_\omega(n))$ converges, it does not converge to the variance of the limiting distribution, which is 0. In this example, the limit of the variances is not the variance of the limit.

Now, it is not that our method is merely insufficient to prove localization for this sequence of potentials; in fact, this example yields a spectrally delocalized operator. Letting $A_n = \cparen{\omega : V_\omega (n) = \frac{1}{n^2}}$ and applying Borel-Cantelli to this family of sets, we see that $V_\omega(n) \neq 0$ for at most finitely many $n$, almost surely. Hence $H_\omega$ is almost surely the free Laplacian plus a compact operator. Since the spectrum of the free Laplacian is $[-2,2]$ and the essential spectrum is invariant under the addition of a compact operator, $H_\omega$ does not enjoy spectral localization in this example.
\end{example}

\section{Example of Sequence Covered by This Result}\label{example_covered}

We wish to present an example of a novel sequence of distributions which is shown to experience localization by this paper. To the author's knowledge, the localization in this example is not implied by any previous results on Anderson localization.

\begin{example} Consider the sequence of potentials defined by
\[V_\omega(n) = \begin{cases}
a_n & \text{with probability } p_n\\
b_n & \text{with probability } 1 - p_n - \epsilon_n\\
\epsilon_n^{-1/\gamma} & \text{with probability } \epsilon_n\\
\end{cases}
\]
where $a_n$ and $b_n$ are any bounded sequences with $|a_n - b_n|$ uniformly bounded away from 0, $\epsilon_n \in [0, 1)$ is any sequence such that $\epsilon_n \to 0$, and $p_n \in (0,1 -\epsilon_n)$ is any sequence bounded away from 0 and 1.

Let $M$ be such that $|a_n|,|b_n| \leq M$ for all $n$. We compute that
\begin{align*}
	\E[|V_\omega(n)|^\gamma] & = |a_n|^\gamma \cdot p_n + |b_n|^\gamma \cdot (1 - p_n - \epsilon_n) + |\epsilon_n^{-1/\gamma}|^\gamma \cdot \epsilon_n\\
	& = |a_n|^\gamma \cdot p_n + |b_n|^\gamma \cdot (1 - p_n - \epsilon_n) + 1\\
	& \leq 2M^\gamma + 1
\end{align*}
for all $n$. Hence we have a uniform bound on the $\gamma$-moment. To check the variance, we will make use of the identity $\mathrm{Var}(X) = \frac{1}{2}\E[(X - X')^2]$ where $X'$ is an independent copy of $X$ (indeed, $\E[(X - X')^2] = \E[X^2] -2\E[XX'] + \E[(X')^2] = 2\E[X^2] -2\E[X]^2 = 2\mathrm{Var}(X)$). Thus, for $n$ sufficiently large, $\epsilon_n^{-1/\gamma} > M$ and so
\begin{align*}
	& \mathrm{Var}(\truncated[M]{V_\omega(n)}) \\
	& = \frac{1}{2}\E\bracket{\paren{\truncated[M]{V_\omega(n)}  - \truncated[M]{V'_\omega(n)}}^2}\\
	& = \frac{1}{2}\left((a_n - b_n)^2p_n(1-p_n - \epsilon_n) + (b_n - a_n)^2p_n(1-p_n - \epsilon_n)\right.\\
	& \quad \left. + (a_n - M)^2\cdot p_n\cdot \epsilon_n + (M- a_n)^2\cdot p_n\cdot \epsilon_n\right)\\
	& = (a_n - b_n)^2p_n(1-p_n - \epsilon_n) + (M- a_n)^2\cdot p_n\cdot \epsilon_n\\
	& \geq (a_n - b_n)^2p_n(1-p_n - \epsilon_n).
\end{align*}
Notice that this is uniformly bounded away from 0 since $|a_n - b_n|$, $p_n$, and $1-p_n - \epsilon_n$ are all uniformly bounded away from 0 by our set-up. Hence the assumptions in \autoref{main_result} are satisfied (regardless of the desired $\gamma$-regime) and we can conclude both spectral and dynamical localization. 
\end{example}


\section*{Acknowledgements}
I would like to thank Anton Gorodetski for presenting me the problem and spending time discussing various issues with me. Thank you to Grigorii Monakov for his help reviewing the manuscript. Thank you as well to Omar Hurtado and Victor Kleptsyn for helpful comments.


\bibliographystyle{myalpha}
\bibliography{references}
\nocite{*}

\end{document}